\documentclass[12pt,a4paper]{article} 
\usepackage{geometry}
\usepackage[english]{babel}
\usepackage{amsmath}
\usepackage{booktabs}
\usepackage{graphicx} 
\usepackage{amssymb} 
\usepackage[hyperfootnotes=false]{hyperref}
\usepackage{xr-hyper} 
\usepackage{bookmark}
\hypersetup{colorlinks = true, citecolor=black, linkcolor=black}
\usepackage{amsthm}
\usepackage{MJOARTI}
\usepackage{setspace} 
\usepackage{rotating}
\usepackage{bbold}
\usepackage{lscape}
\usepackage{subfigure}
\usepackage{caption}
\usepackage{subcaption}
\usepackage{hyperref} 
\usepackage{lscape}
\usepackage{ifthen}
\usepackage{url}
\usepackage{blkarray}
\usepackage{multirow}
\usepackage{tabularx}
\usepackage{longtable}
\usepackage{enumerate}
\usepackage{supertabular}
\usepackage{fancyhdr}
\usepackage{epstopdf}
\usepackage{dsfont}
\usepackage{url}
\usepackage{multicol}
\usepackage{titlesec}
\usepackage{eurosym}
\usepackage{arydshln}
\usepackage{natbib}
\usepackage{threeparttable} 
\usepackage{comment}
\usepackage{xcolor}
\usepackage{pdflscape} 
\usepackage{csquotes}
\usepackage{placeins}
\usepackage{needspace} 
\usepackage{amsfonts} 
\usepackage{adjustbox} 

\usepackage{hyperref}
\usepackage{xr-hyper}

\usepackage{longtable}
\usepackage{booktabs}
\usepackage{threeparttablex}

\usepackage{bibunits}

\usepackage{array}
\newcolumntype{k}{>{\centering\arraybackslash}p{2cm}}
\usepackage{longtable}
\usepackage{ltcaption}
\usepackage{multicol}
\usepackage{color}
\usepackage{textcomp}
\definecolor{dark-red}{rgb}{0.6,0,0}
\definecolor{listinggray}{gray}{0.9}
\definecolor{darkgreen}{rgb}{0,0.4,0}
\definecolor{lbcolor}{rgb}{0.9,0.9,0.9}

\begin{document}
 
\title{
\textbf{Publication bias and $p$-hacking in the effect of COVID-19 on learning}%
\thanks{
Authors: Luskova: Institute of Economic Studies, Faculty of Social Sciences, Charles University. E-mail: \href{mailto:martina.luskova@fsv.cuni.cz}{martina.luskova@fsv.cuni.cz}. Buliskeria, corresponding author: Department of Economics, Nazarbayev University. E-mail: \href{mailto:nino.buliskeria@nu.edu.kz}{nino.buliskeria@nu.edu.kz}. Elminejad: Department of Economics, Nazarbayev University. E-mail: \href{mailto:ali.elminejad@nu.edu.kz}{ali.elminejad@nu.edu.kz}. Havranek: Institute of Economic Studies, Faculty of Social Sciences, Charles University; also affiliated with METRICS, Stanford University, and CEPR, London. E-mail: \href{mailto:tomas.havranek@fsv.cuni.cz}{tomas.havranek@fsv.cuni.cz}. Irsova: Institute of Economic Studies, Faculty of Social Sciences, Charles University; also affiliated with METRICS, Stanford University, and Anglo-American University, Prague. E-mail: \href{mailto:zuzana.irsova@fsv.cuni.cz}{zuzana.irsova@fsv.cuni.cz}. Jurajda: CERGE-EI, a joint workplace of the Center for Economic Research and Graduate Education of Charles University and the Economics Institute of the Czech Academy of Sciences. E-mail: \href{mailto:stepan.jurajda@cerge-ei.cz}{stepan.jurajda@cerge-ei.cz}. Kapicka: CERGE-EI, a joint workplace of the Center for Economic Research and Graduate Education of Charles University and the Economics Institute of the Czech Academy of Sciences. E-mail: \href{mailto:marek.kapicka@cerge-ei.cz}{marek.kapicka@cerge-ei.cz}. The authors declare no conflicts of interest related to this work. \\ 
\indent Acknowledgment: We thank Bastian A. Betthäuser, Anders M. Bach-Mortensen, and Per Engzell for helpful comments. Tomas Havranek and Zuzana Irsova acknowledge support from the Czech Science Foundation, project 24-11583S. Martina Luskova acknowledges support from the Charles University Grant Agency, project No. 136724. 
}}

\author{Martina Luskova, Nino Buliskeria, Ali Elminejad, Tomas Havranek, \\Zuzana Irsova, Stepan Jurajda, Marek Kapicka}

\date{June 13, 2026}

\maketitle

\renewcommand{\abstractname}{Abstract}

\begin{abstract}
\bigskip

We revisit a central estimate in the economics of education: the human-capital loss associated with COVID-19 school closures. Estimates of pandemic learning loss may be affected by publication bias, $p$-hacking, and the mechanical correlation between standardized effect sizes and their standard errors. We conduct a comprehensive multi-method assessment of bias by applying a wide range of correction techniques -- including PET-PEESE, three-parameter selection models (3PSM), Robust Bayesian Meta-Analysis (RoBMA), Meta-Analysis Instrumental Variable Estimation (MAIVE), Right-Truncated Meta-Analysis (RTMA), and multi-bias sensitivity analysis. Our preferred specifications, RoBMA and MAIVE, rely on different assumptions yet converge on an effect size of approximately $-0.12$ SD, equivalent to a learning loss of about 30\% of a school year. Although some methods reveal signs of publication bias and selective reporting, these findings do not explain away the central finding: the COVID-19 learning deficit is economically meaningful and statistically robust.  


\end{abstract}
 
\clearpage
\doublespacing
\clearpage
\section{Introduction}

The COVID-19 pandemic triggered an unprecedented disruption to formal education worldwide. School closures, shifts to remote learning, and broader social disruptions affected hundreds of millions of students across all income levels and regions. As schools reopened and assessment data accumulated, a large empirical literature emerged documenting what is now widely referred to as ``pandemic learning loss'' — declines in student achievement relative to pre-pandemic benchmarks or counterfactual trends \citep{unitednations2020,engzell2021learning, donnelly2022learning}. Understanding the magnitude of these losses directly informs remediation policy design, public expenditure decisions, and the longer-run assessment of how major systemic shocks accumulate into human-capital deficits.

For economists, the magnitude of pandemic learning loss matters because test-score deficits are widely used as proxies for losses in human-capital accumulation, with implications for inequality, remediation spending, and long-run productivity \citep{hanushek2008role,hanushek2020economic}. The policy question is therefore not only whether students learned less during the pandemic, but whether the estimated magnitude is robust enough to guide large-scale catch-up interventions. Answering this question requires assessing whether the reported effect sizes are robust to publication bias, selective reporting, and the statistical complications introduced by standardized effect-size measures. 
 
Several meta-analyses have quantified the extent of pandemic learning loss \citep{konig2022impact, betthauser2023systematic, di2023impact, wisenocker2025meta}. These syntheses report average deficits of roughly $-0.14$ to $-0.20$ standard deviations (Cohen's $d$). Where moderator analyses are available, losses tend to be larger in mathematics, and several studies also report larger losses among disadvantaged students. We take as our empirical benchmark the most prominent and widely cited synthesis, \citet{betthauser2023systematic}, which conducts a meta-analysis of 291 effect-size estimates from 42 studies across 15 countries and reports an average learning loss of approximately $-0.14$ standard deviations (Cohen's $d$). Following \citet{betthauser2023systematic}, we interpret this magnitude using the common benchmark that students typically gain about $0.40$ standard deviations per school year under normal circumstances \citep{azevedo2021simulating,bloom2008performance,hill2008empirical} -- this corresponds to roughly $35\%$ of a school year. Existing assessments of publication bias in this literature, including \citet{betthauser2023systematic}, have relied primarily on visual or conventional diagnostic tools, including funnel plots, Egger's test, Doi plots, distributions of $z$-statistics, and $p$-curves. These assessments generally conclude that publication bias is not a major concern.

Although these diagnostics are widely used and provide useful first-pass evidence, they remain limited in their ability to model selection directly or to quantify how sensitive the pooled estimate is to selective reporting. Moreover, expressing estimates as Cohen's $d$ can mechanically induce a correlation between standardized effect sizes and their standard errors. This issue is not specific to any individual study, but arises from the construction of standardized effect-size measures, which is standard practice across the social sciences. We therefore extend the existing assessments by applying regression-based, selection-model, instrumental-variable, and sensitivity-analysis approaches that provide a more formal evaluation of publication bias, $p$-hacking, and their implications for the estimated magnitude of pandemic learning loss.

This paper provides a systematic econometric assessment of bias in the COVID-19 learning-loss literature. Using the dataset of \citet{betthauser2023systematic}, we apply a wide range of state-of-the-art bias-detection and correction techniques: the Precision Effect Test and Precision Effect Estimate with Standard Errors (PET-PEESE), the Three-Parameter Selection Model (3PSM), Robust Bayesian Meta-Analysis (RoBMA), Meta-Analysis Instrumental Variable Estimation (MAIVE), Right-Truncated Meta-Analysis (RTMA), and multi-bias sensitivity analysis. These methods differ in their assumptions, their sensitivity to heterogeneity, and their ability to distinguish $p$-hacking from publication bias. Crucially, RoBMA and MAIVE are less reliant on funnel-based assumptions in settings where standardized effect sizes, such as Cohen's $d$, mechanically induce dependence between effect-size estimates and their standard errors: MAIVE addresses this dependence directly by instrumenting precision, while RoBMA avoids committing to the funnel-asymmetry slope that this dependence distorts. The paper therefore contributes to the economics literature in two ways. Substantively, it reassesses a central estimate in the economics of education: the human-capital loss associated with COVID-19 school closures. Methodologically, it shows how publication-bias, $p$-hacking, and endogenous-precision corrections behave in a high-powered applied literature based on standardized effect sizes. Because we use the same dataset as \citet{betthauser2023systematic}, computationally reproduce their main results --- as documented in Appendix~\ref{app:reproduction} --- and reanalyze the same research question using additional bias-correction methods, the paper also constitutes a reproducibility and replication exercise in the spirit of the meta-science literature.

Our findings support the original conclusion, but on broader methodological grounds. Across correction methods, bias-adjusted estimates consistently point to a meaningful decline in student achievement. Our preferred estimates from RoBMA and MAIVE converge on an effect size of approximately $-0.12$, equivalent to a learning loss of roughly 30\% of a school year. Although we find evidence consistent with selective reporting that is not apparent from graphical diagnostics alone, these biases do not overturn the central finding: COVID-19 learning losses are robust and substantively meaningful. The modest downward revision from approximately $-0.14$ to $-0.12$ does not materially alter the policy implications of the original estimate: a deficit of this magnitude still represents a non-trivial and potentially persistent loss in human-capital accumulation for the affected cohorts.

The remainder of the paper is organized as follows. Section~\ref{sec:data} describes the dataset, summarizes key patterns of heterogeneity, and presents descriptive diagnostics that motivate the formal bias analysis. Section~\ref{sec:pubbias} applies publication-bias correction and sensitivity methods, including PET-PEESE, 3PSM, RoBMA, MAN, selection-ratio adjustments, and $s$-values. Section~\ref{sec:phacking} turns to methods that additionally address within-study selection, endogenous precision, and multiple sources of bias, including MAIVE, RTMA, and multi-bias sensitivity analysis. Section~\ref{sec:conclusion} concludes by summarizing the evidence and discussing its implications for the robustness of estimated COVID-19 learning losses. 

\section{Background and Data}
\label{sec:data}

Our analysis builds on the dataset compiled by \citet{betthauser2023systematic}. 
We use the same set of studies and effect-size estimates, so any difference between our findings and theirs reflects differences in aggregation and bias-correction methods rather than differences in study selection or effect-size construction. The dataset is publicly available via the Open Science Framework \citep{betthauser2023dataset} at
\url{https://doi.org/10.17605/osf.io/u8gaz}. Replication code for all analyses reported in this  paper is available at \url{https://meta-analysis.cz/learning}. Throughout the analysis, we follow recent guidance for meta-analysis in economics provided by \citet{irsova2024meta} and \citet{cook2026guidance}, as well as the reporting standards proposed by \citet{havranek2020reporting} and \citet{cook2026reporting}. Appendix~\ref{app:guidelines} documents how the study addresses these recommendations item by item. 

The dataset contains 291 effect-size estimates drawn from 42 studies across 15 countries (Figure~\ref{fig:country}). It is geographically concentrated and dominated by large administrative records: roughly half of all estimates come from the United States, while the United Kingdom and the Netherlands account for most of the remainder. Study sizes vary substantially, from entire national school populations numbering in the millions to samples of only a few hundred students.

The data reveal substantial heterogeneity across key dimensions (Tables~\ref{tab:dscrstat} and~\ref{tab:subgroups}): mathematics estimates are markedly larger in magnitude than reading estimates, consistent with evidence from other educational disruptions that mathematics learning depends more heavily on formal instruction (Figure~\ref{fig:heterogeneity}, panel a). By contrast, there is little evidence of a systematic grade-level gradient: primary and secondary school estimates are nearly identical on average (Figure~\ref{fig:heterogeneity}, panel b), and a finer-grained breakdown by individual grade confirms that effect sizes remain close to the overall mean across grades 1--9 (Figure~\ref{fig:grade}). The more negative estimates in grades 11--13 should be interpreted cautiously because they are based on very few observations, and the confidence band is correspondingly wide. Each study receives equal total weight in the subgroup averages, so that prolific studies, i.e., studies producing multiple effect estimates, do not dominate the pooled estimates, and study-level clustering accounts for the resulting within-study dependence throughout. Our corpus spans peer-reviewed articles, working papers, and institutional and commercial reports (Table~\ref{tab:subgroups}); the relevant selection therefore operates across reporting outlets rather than through journal acceptance alone, and we use ``publication bias'' throughout in the broad sense of selective \emph{availability} of results.

The estimates in the dataset are all expressed as Cohen's $d$, which is necessary to aggregate results across studies using different assessment instruments. This standardization has an important methodological consequence, to which we return in Section~\ref{sec:pubbias}: because both the standardized effect size and its standard error are functions of the same underlying standard deviation, they may be mechanically correlated across studies. This dependence complicates the interpretation of funnel-plot-based publication-bias diagnostics, which rely precisely on the relationship between reported effects and their precision.

We begin with descriptive diagnostics of the structure of the data. Figure~\ref{fig:funnel} presents a conventional funnel plot with effect sizes on the horizontal axis and estimated standard errors on the vertical axis. The dotted vertical line marks the pooled mean effect, while the dashed vertical line marks zero. Most estimates are negative and lie to the left of zero, consistent with an overall learning deficit. The most precise estimates, located near the bottom of the figure, cluster around modestly negative effects, while less precise estimates display substantially greater dispersion. The cluster of highly precise estimates near $SE \approx 0$ reflects the presence of very large studies in the dataset. Figure~\ref{fig:significantvsnot} separates statistically significant and non-significant estimates: significant estimates extend farther into the left tail, whereas non-significant estimates cluster closer to zero. Figure~\ref{fig:sig_funnel} presents the corresponding significance funnel, which we examine in Section~\ref{sec:pubbias}. This visual pattern is consistent with the possibility of selective reporting favoring negative and statistically significant learning-loss estimates, but it does not provide a clean visual basis for distinguishing publication bias from genuine heterogeneity or from mechanical dependence introduced by standardizing estimates as Cohen's $d$. It instead motivates the formal bias-correction and sensitivity analyses reported below. 

We therefore also examine threshold-based diagnostics. Figures~\ref{fig:diagnostic_z} and \ref{fig:comparison} plot the distribution of $z$-statistics. The log-transformed distribution does not show clear clustering around the conventional $1.96$ significance threshold, consistent with \citet{betthauser2023systematic}. In the raw distribution, however, the $z$-statistics show visible mass near the negative 5\% significance threshold and a long left tail. This pattern is consistent with a preference for negative statistically significant estimates, but it is not dispositive: bunching near critical values can also arise when many studies have similar power or sample sizes \citep{elliott2022power}.

We additionally apply a caliper test and attempt a $p$-curve analysis to assess whether the data show evidence of local threshold manipulation. The caliper test finds no robust evidence of excess mass just inside the negative 5\% significance threshold, $z=-1.96$, with only weak and non-monotonic significance at wider calipers. The $p$-curve is also not informative in this setting: the distribution is highly compressed near zero because the literature includes very large studies, with a median sample size of $n=50{,}000$ and a maximum sample size of $n=10{,}884{,}922$. These sample sizes generate extremely small standard errors, so even moderate effect sizes can produce very large $z$-statistics. Full results are reported in Appendix~\ref{app:pcaliper}. Thus, the threshold diagnostics provide little evidence of local threshold manipulation, even though some broader features of the distribution remain consistent with selective reporting.

As a further robustness check, we examine case-level influence using DFBETAS,  which measures how much the pooled estimate changes when each observation is  omitted in turn.\footnote{Influence diagnostics require a standard random-effects model; we use the REML estimate ($\hat{\mu} = -0.124$).} Under the sample-size-adjusted threshold  $|\text{DFBETAS}| > 2/\sqrt{n} = 0.117$, eleven estimates are flagged: nine  with negative DFBETAS values (range: $-0.28$ to $-0.12$), pulling the pooled  estimate in the negative direction, and two with positive values ($+0.14$ and  $+0.15$), pulling in the opposite direction. The net effect is negligible:  excluding all eleven moves the pooled estimate from $-0.124$ to $-0.116$,  confirming that no individual estimate materially drives the main result.  The two most influential estimates --- those from \citet{ardington2021covid}  ($\text{DFBETAS} = -0.28$ and $-0.22$) --- report learning losses from  South Africa, one of the few middle-income countries in the dataset and among  those with the most severe school disruptions during the pandemic. Their  influence is consistent with genuine heterogeneity rather than measurement  error. The DFBETAS plot is presented in Figure~\ref{fig:dfbetas}.

Taken together, these figures motivate rather than settle the bias question. The estimates are predominantly negative, and some descriptive patterns are consistent with selective reporting, but visual and threshold-based diagnostics alone are insufficient in this setting. The following sections therefore apply regression-based, selection-model, instrumental-variable, right-truncated, and sensitivity-analysis methods that more directly evaluate publication bias and $p$-hacking.

\section{Correcting for publication bias}
\label{sec:pubbias}

As a benchmark, Panel~A of \autoref{tab:pubbias} presents the standard  uncorrected point estimates: an equal-weighted (unweighted) mean of $-0.126$  ($SE = 0.059$) and a robust variance estimation (RVE) estimate of $-0.140$  ($SE = 0.020$) \citep{hedges2010robust}, which accounts for within-study  dependence by clustering at the study level. These estimates align with  \citet{betthauser2023systematic} precisely because they represent the  uncorrected mean, assuming no selection bias. All correction methods below  are assessed relative to this uncorrected mean of $-0.140$. 

We begin with the funnel-based precision effect test (PET-PEESE). PET is a weighted regression of the effect on its standard error, with weights $1/SE^2$; the significant coefficient on the standard error is consistent with funnel asymmetry (\autoref{tab:pubbias}, Panel~B, Column~1). PEESE regresses the effect on the squared standard error and provides a more accurate bias-corrected estimate \citep{stanley2017limitations, bartovs2022adjusting}, yielding a corrected estimate of $-0.245$, corresponding to $0.245/0.4 = 0.61$ school years of learning deficit (\autoref{tab:pubbias}, Panel~B, Column~2). However, as noted in Section~\ref{sec:data}, the standardization to Cohen's $d$ mechanically links estimates and standard errors across studies, which means funnel-based results should be interpreted cautiously. We address this directly in Section~\ref{sec:phacking}.

A selection-model approach yields a more moderate correction. The three-parameter selection model (3PSM) applies maximum likelihood to correct for the preferential publication of $p$-values below the significance threshold \citep{iyengar1988selection, hedges1992selectionmodeling, vevea1995general}. The corrected estimate is $-0.123$, equivalent to $0.123/0.4 = 0.31$ school years of learning loss. The associated likelihood-ratio test does not reject the null hypothesis of no publication bias (\autoref{tab:pubbias}, Panel~B, Column~3). Relative to PEESE, this estimate is considerably closer to the uncorrected mean of $-0.140$, suggesting that conclusions about the magnitude of bias depend materially on the correction method.

Finally, we apply Robust Bayesian Meta-Analysis (RoBMA), which constructs a model-averaged estimate across multiple publication bias models, weighting by data fit and parsimony \citep{maier2023robustBMA, bartovs2023robustBMA}. The RoBMA estimate is $-0.118$, or $0.118/0.4 = 0.30$ school years (\autoref{tab:pubbias}, Panel~B, Column~4). Because RoBMA integrates over uncertainty about the appropriate bias model rather than committing to one, it is robust to misspecification and performs well under heterogeneity. We treat this as our preferred bias-corrected estimate from this section. The sensitivity of this estimate to increasingly severe forms of selective publication is assessed in the following paragraphs, and further corroborated by regression-based correction methods robust to $p$-hacking in Section~\ref{sec:phacking}.

\textit{\textbf{Sensitivity to selective publication}}. The correction methods in Table~\ref{tab:pubbias} each rely on a specific model of the selection process. We complement them with a sensitivity-based approach that asks how robust the findings are to a range of assumed degrees of selective publication, following \citet{mathur2020sensitivity, mathur2024assessing, mathur2024p}. We implement these analyses using the \texttt{PublicationBias} and \texttt{phacking} packages available at \href{https://metabias.io}{metabias.io}. These methods assess how sensitive conclusions are to the selective reporting of affirmative results --- those that are statistically significant in the expected direction --- relative to non-affirmative results, which include non-significant or unexpected-direction estimates. The degree of this asymmetry is quantified by the \textit{selection ratio}: the relative likelihood of an affirmative result being published compared to a non-affirmative one. A selection ratio of one indicates no preference for significant results. In our setting, an affirmative result is one that is negative and statistically significant, since learning deficit is measured as a negative value.\footnote{In Figures~\ref{fig:forest_replication} through~\ref{fig: variations} and \autoref{fig:sig_funnel}, most coefficient estimates are negative, as expected. We account for the negative sign and adjust model specifications accordingly, assuming one-directional selection that favors negative, statistically significant estimates, using a two-sided significance threshold of $\alpha = 0.05$ (i.e.\ $z < -1.96$).} Because the true degree of publication bias is unknown, we assess robustness across a range of assumed selection ratios rather than committing to one value.  

\textit{\textbf{Meta-analysis of non-affirmative results (MAN)}}. In the extreme case where affirmative studies are infinitely more likely to be published than non-affirmative ones, the worst-case bias-corrected estimate is obtained from a meta-analysis restricted to non-affirmative results only. As \citet{mathur2024assessing} notes, MAN does not measure the actual strength of publication bias; rather, it is a stress test that assesses how results hold up under the most extreme possible form of selective publication. Since it relies only on non-affirmative results, it is typically biased toward the null when affirmative results are selectively favored.\footnote{\citet{mathur2024p} shows that MAN remains conservative under $p$-hacking when $p$-hacking favors affirmative outcomes, because it places no weight on affirmative estimates, which are the estimates favored by such selection.} In our setting, MAN asks whether the negative learning-loss effect survives after placing all weight on the least favorable subset of estimates, and is therefore best interpreted as a deliberately conservative robustness check rather than as a preferred corrected estimate. If the worst-case estimate still aligns with the uncorrected mean in sign and magnitude, this provides strong evidence of robustness. Panel~A, Column~1 of \autoref{tab:phack} reports a MAN estimate of $0.021$, which is statistically significant at the 10\% level but not at the 5\% level. This estimate reverses sign relative to the uncorrected random-effects mean of $-0.140$, suggesting that the results are sensitive to worst-case selective publication and motivating the examination of less extreme scenarios.

\textbf{\textit{Selection ratio sensitivity.}} Following \citet{mathur2020sensitivity, mathur2024p}, we use a four-fold selection ratio as an illustrative benchmark, where affirmative studies are assumed to be four times more likely to be published than non-affirmative ones. Under this assumption, both the fixed- and random-effects estimates retain the negative sign of the uncorrected mean (\autoref{tab:phack}, Panel~A, Columns~2 and~3). The fixed-effects estimate is larger in magnitude, at $-0.206$, while the robust random-effects estimate is smaller, at $-0.068$, reflecting its adjustment for heterogeneity and clustering of estimates within studies. Thus, even under a relatively strong assumed degree of selective publication, the corrected estimates continue to indicate learning losses, although their magnitude depends on how heterogeneity and within-study clustering are handled. The range between these estimates also contains our preferred bias-corrected estimates from RoBMA ($-0.118$) and MAIVE ($-0.119$, reported in Section~\ref{sec:phacking}), suggesting that those estimates are consistent with a moderate rather than worst-case degree of selective publication.

\textit{\textbf{s-Value.}} We next invert the sensitivity exercise and ask how strong selective publication would have to be to explain away the result. The $s$-value is the selection ratio required to shift the estimated effect or its confidence interval bound to a specified value. Small $s$-values indicate that a relatively modest degree of selective publication could explain away the result, whereas large $s$-values indicate greater robustness. In our case, affirmative studies would need to be $29.60$ times more likely to be published than non-affirmative studies to shift the point estimate to zero, and $8.31$ times more likely to shift the confidence interval bound to zero (\autoref{tab:phack}, Panel~B, Column~1). Moreover, no degree of selection bias under this model can shift either the point estimate or the confidence interval bound to $+0.05$ (\autoref{tab:phack}, Panel~B, Column~3). These results suggest that, although the MAN estimate indicates sensitivity to the most extreme form of selective publication, the main directional conclusion would require a very large degree of selective publication to be overturned and is therefore robust to plausible degrees of publication bias. 

\textit{\textbf{Significance funnel plot.}} As a supplement to the selection-ratio sensitivity analysis, we present the significance funnel plot in \autoref{fig:sig_funnel}. The figure separates affirmative estimates (orange points) from non-affirmative estimates (gray points) and visualizes the extent to which the mean estimate depends on affirmative results. The contrast between the two groups is substantial.   The gray diamond shows the mean among non-affirmative estimates ($0.021$),  corresponding to the MAN estimate reported in \autoref{tab:phack}, Panel~A,  Column~1. The black diamond\footnote{\label{fn:uwls} The black diamond corresponds to the inverse-variance weighted  mean from \citet{mathur2020sensitivity}, estimated as a fixed-effects  model ($\hat{\mu}_{IVW} = -0.245$, SE $= 0.0001$). The unrestricted  weighted least squares (UWLS) estimator \citep{stanley2023unrestricted, stanley2017neither},  which uses the same inverse-variance weights but does not impose  distributional assumptions on the residuals, yields an identical point  estimate with a more conservative cluster-robust standard error  (SE $= 0.051$, $p < 0.001$). The large discrepancy in standard errors  reflects the high degree of heterogeneity in the dataset  ($I^2 = 99.97\%$), which the fixed-effects model ignores but the  cluster-robust UWLS standard error partially accounts for. Both  estimates are statistically significant; we report the fixed-effects  version following \citet{mathur2020sensitivity} and note the UWLS  equivalence in the interest of transparency.}  shows the precision-weighted (inverse-variance)  pooled mean across all 291 estimates ($\hat{\mu}_{IVW} = -0.245$), which lies  more negative than the equal-weighted mean of $-0.126$ reported in Panel~A of  Table~\ref{tab:pubbias} because inverse-variance weighting gives disproportionate  influence to large, high-precision studies that tend to report larger learning  losses.  The gap between the black and gray diamonds is consistent with some degree of selective publication and suggests that the magnitude of the pooled effect may be exaggerated \citep{mathur2020sensitivity}. At the same time, the non-affirmative mean should not be interpreted as the most credible corrected estimate of the underlying effect. It is the visual counterpart to the MAN worst-case benchmark. The figure therefore reinforces the MAN result: the evidence is sensitive to worst-case selective publication, but this stress test is intentionally severe.

Taken together, the sensitivity analyses suggest that the results are sensitive to the most extreme form of selective publication, but that overturning the negative directional finding would require a very large degree of bias. The MAN estimate therefore serves as a severe stress test rather than as our preferred corrected estimate. By contrast, the preferred RoBMA and MAIVE estimates, reported above and in Section ~\ref{sec:phacking}, respectively, are consistent with the range implied by finite selection-ratio adjustments, suggesting that the main conclusion is robust to plausible degrees of publication bias. 
 
\section{Accounting for \textit{p}-hacking and multiple biases}
\label{sec:phacking} 
 
The preceding section evaluates sensitivity to selective publication across studies. We now consider whether the conclusion changes once we account for biases that may arise within studies before estimates enter the meta-analysis. Following \citet{mathur2024p}, we distinguish publication bias, or selection across studies, from \textit{p}-hacking, or selection within studies. In practice, \textit{p}-hacking may involve searching across specifications, samples, covariate sets, or outcomes until an affirmative result is obtained. Because such decisions can affect both the measured underlying quantity and the reported precision, \textit{p}-hacking can distort the primary-study estimate itself, rather than simply determining which otherwise unbiased estimate is written up or published.\footnote{For detailed discussions of \textit{p}-hacking, see \cite{brodeur2020methods, elliott2022detecting, brodeur2023unpacking, mathur2020sensitivity, mathur2024p}.}

These concerns matter because conventional publication-bias corrections typically assume that primary-study estimates are, individually, unbiased for their corresponding study-specific population effects. In that framework, bias enters through the selection mechanism: some otherwise unbiased estimates are more likely than others to be written up, submitted, or accepted for publication. When \textit{p}-hacking or spurious precision is present, however, the estimates, their reported standard errors, or both may already be distorted by within-study research choices.

We therefore proceed in three steps. First, we apply MAIVE, which instruments for potentially endogenous precision and is especially relevant when standardized effect sizes mechanically link estimates and standard errors (e.g., Cohen's $d$). Second, we apply RTMA, which directly models \textit{p}-hacking as selection within studies. Third, we apply the multi-bias framework of \citet{mathur2024internal}, which allows publication bias to interact with internal study bias by allowing affirmative and non-affirmative estimates to differ in their probability of publication and in their average degree of internal bias.

\textit{\textbf{MAIVE.}} We first apply the Meta-Analysis Instrumental Variable Estimator (MAIVE) of \citet{irsova2025MAIVE}. MAIVE is designed for settings in which reported standard errors may themselves be endogenous. It has increasingly been used in recent meta-analyses in economics to address spurious precision, including applications in labor economics, education, and microeconomics \citep{havranek2024publication,opatrny2025class,cala2026incentives}.
Conventional funnel-based methods, such as PET and PEESE, treat precision as observed and place greater weight on estimates with smaller standard errors. In empirical research, however, precision is estimated by researchers and may be affected by specification choices, sample restrictions, or other research decisions. If these choices affect both the reported coefficient and its standard error, then the standard error is endogenous, and methods relying on the relationship between estimates and standard errors may themselves become biased \citep{brodeur2023unpacking, irsova2025MAIVE, mathur2024p}.

MAIVE addresses this problem by instrumenting for the reported variance. Following \citet{irsova2025MAIVE}, we use inverse sample size as an instrument for the squared standard error. The relevance condition is motivated by the mechanical relationship between sampling variance and sample size: all else equal, larger samples imply lower sampling variance. The exclusion restriction is motivated by the fact that sample size is generally harder to manipulate ex post than reported precision. Researchers may be able to affect standard errors through specification choices, sample restrictions, clustering decisions, or alternative outcome definitions, but the underlying sample size is often determined before estimation, especially in observational and administrative-data settings. Sample size is also directly observed and does not suffer from measurement error, whereas the reported standard error is itself an estimate and may vary with methodological choices. A possible limitation is that sample size may still be endogenous if researchers who anticipate smaller effects design larger studies. In our setting, however, many estimates come from observational or administrative data in which sample size is largely predetermined.

This approach is particularly relevant here because all estimates are standardized as Cohen's $d$. Such standardization can mechanically induce a correlation between effect sizes and their standard errors, violating the independence condition that underlies standard funnel-based meta-analytic models in the absence of publication bias. Applying MAIVE, we therefore regress the squared reported standard errors on inverse sample size and use the predicted values in place of the reported variance in the meta-regression. For our baseline MAIVE specification, we use the instrumented version of PET-PEESE without additional inverse-variance weighting and cluster standard errors at the study level, as recommended by \citet{irsova2025MAIVE}.
 
The MAIVE estimate, reported in Column~1 of Panel~C in \autoref{tab:phack}, is $-0.119$ and statistically significant. The first stage confirms instrument relevance: inverse sample size strongly predicts reported squared standard errors ($F = 271.716$). The Anderson--Rubin 95\% confidence interval, $[-0.137, -0.087]$, remains entirely below zero. The estimated publication-bias component (the coefficient on the fitted $SE^2$) is $-0.245$ and statistically significant at the 1\% level, suggesting that publication bias remains detectable even after instrumenting for the potentially endogenous relationship between reported effects and precision. Substantively, the MAIVE estimate corresponds to a learning deficit of $0.119/0.4 = 0.30$ school years. This estimate is nearly identical to the RoBMA estimate of $-0.118$ reported in \autoref{tab:pubbias} and close to the uncorrected random-effects mean of $-0.140$. Because MAIVE directly addresses endogenous precision and the mechanical correlation introduced by standardization to Cohen's $d$, we treat it, together with RoBMA, as one of the most credible estimates in our analysis. Taken together, these estimates point to a learning deficit of approximately 30\% of a school year.

\textit{\textbf{RTMA.}} While MAIVE addresses endogenous precision that may arise from \textit{p}-hacking or from the standardization to Cohen's $d$, it does not explicitly model \textit{p}-hacking as a within-study search process. We therefore next apply the right-truncated meta-analysis (RTMA)\footnote{using \texttt{PublicationBias} and \texttt{phacking} packages available at \href{https://metabias.io}{metabias.io}.} of \citet{mathur2024p}, which is designed to account for \textit{p}-hacking as selection within studies, while also allowing for selective publication across studies. RTMA is correctly specified when the favored estimates in \textit{p}-hacked studies are affirmative, meaning that researchers search across estimates until an affirmative result is obtained, or when \textit{p}-hacked studies with non-affirmative favored estimates are not published. The method also accounts for within-study heterogeneity, allowing both independent and autocorrelated estimates within studies and providing a framework for inference. The resulting RTMA estimate,\footnote{In our application, affirmative results are negative and statistically significant estimates of learning loss. Since the RTMA implementation is formulated for settings in which affirmative results are positive and statistically significant, we apply the method to sign-reversed estimates and then reverse the sign of the resulting estimate. See Appendix~\ref{app:rtma} for a detailed discussion of this implementation choice and a related coding issue in \texttt{phacking\_meta}.} reported in Column~2 of Panel~C in \autoref{tab:phack}, is $-0.039$, corresponding to approximately $0.039/0.4 = 0.10$ school years of learning loss. The estimate remains negative but is much closer to zero than the MAIVE and RoBMA estimates. 

Because RTMA is formulated as a random-effects model, it does not impose a single common true effect across studies; instead, it estimates an underlying distribution of population effects, summarized by a mean effect and a heterogeneity parameter. The estimated heterogeneity parameter is $0.072$, which captures the estimated standard deviation of the underlying true effects around the corrected mean \citep{borenstein2021introduction, mathur2024p}. Thus, the fitted RTMA model implies a relatively small average learning-loss effect, centered at $-0.039$, together with meaningful dispersion in the underlying true effects. Because RTMA relies on assumptions about the distribution of published non-affirmative estimates and the selection process, we examine its diagnostic fit before interpreting the estimate substantively.

\textit{\textbf{Diagnostic plots for RTMA.}} Following \citet{mathur2024p}, we examine two diagnostic plots -- the $z$-score distribution and the Quantile-Quantile (Q-Q) plot showing the fitted versus the empirical CDF functions. The $z$-score distribution helps assess whether the observed pattern of selective reporting is consistent with the one-directional selection assumed by RTMA. In our setting, affirmative estimates are negative and statistically significant, so concentration around $z=-1.96$ is consistent with selection favoring negative significant learning-loss estimates. By contrast, comparable concentration around both $-1.96$ and $+1.96$, or concentration favoring non-affirmative significant estimates, would raise concerns about violations of RTMA's selection assumptions. The RTMA diagnostic Q-Q plot assesses whether the published non-affirmative estimates are well described by the fitted truncated distribution implied by the RTMA estimates of the mean effect and between-study heterogeneity.  

\autoref{fig:diagnostic_z} displays the distribution of $z$-scores in the original data. The distribution is left-skewed, with visible mass near the negative critical threshold $z=-1.96$, consistent with a preference for negative statistically significant results. At first glance, this pattern could be interpreted as evidence of \textit{p}-hacking. However, \autoref{fig:comparison}~(b) shows mass concentrated near $|z|=1.96$; because it is computed on absolute values, this does not by itself distinguish directional selective reporting from features of the power or sample-size distribution \citep{elliott2022power}. The caliper test in \autoref{tab:caliper} points in the same direction: we find no significant excess of estimates just inside conventional significance thresholds.

\autoref{fig:diagnostic_qq} presents the RTMA diagnostic Q-Q plot, which compares the fitted CDF of the published non-affirmative estimates with their empirical CDF. The points adhere closely to the 45-degree line in the lower quantiles, but deviations become more pronounced toward the right-hand side of the distribution, with dispersion increasing at higher quantiles. This pattern suggests that RTMA captures the lower part of the distribution reasonably well but fits the upper tail less accurately. One possible explanation is that the mechanical relationship between Cohen's $d$ and its standard error distorts the distribution of non-affirmative estimates in ways not fully captured by the truncated normal model, as discussed in Sections~\ref{sec:data} and~\ref{sec:pubbias}.

Taken together, the diagnostics do not provide strong evidence of \textit{p}-hacking, and the imperfect tail fit in the RTMA diagnostic plot suggests that the RTMA point estimate should be interpreted cautiously. We therefore view RTMA as a useful robustness check rather than a preferred correction. 

\textit{\textbf{Multiple biases.}} The preceding analyses separately address publication bias, spurious precision, and \textit{p}-hacking. A remaining concern is that selective publication may interact with internal study bias. For example, if internally biased studies are more likely to produce large negative estimates, and if large negative estimates are more likely to be published, then publication bias may disproportionately select studies with greater internal bias. In this case, correcting only for selective publication, while assuming that all studies have the same average internal bias, may understate the combined distortion.

To address this possibility, we apply the multi-bias sensitivity framework of \citet{mathur2024internal}. The method asks how the meta-analytic estimate would change under specified assumptions about both publication bias and average internal bias.\footnote{(1) \textit{``For a given severity of internal bias across studies and of publication bias, how much could the results change?''}; and (2) \textit{``For a given severity of publication bias, how severe would internal bias have to be, hypothetically, to
attenuate the results to the null or by a given amount?''}}

The framework can allow internal bias to affect all studies or only selected subsets, such as non-randomized studies. Ideally, one would therefore allow internal bias to differ by study design, for example by assigning different bias parameters to randomized and non-randomized studies. Because we do not have complete information on which estimates come from randomized designs, we take a more aggregate approach and allow average internal bias to differ between affirmative and non-affirmative estimates. This specification directly addresses the concern that affirmative estimates may differ from non-affirmative estimates not only in their probability of publication, but also in their average degree of internal bias.

Columns~3 and~4 of Panel~C in \autoref{tab:phack} report the multi-bias results. As a benchmark, we use the robust random-effects selection-ratio estimate with a four-fold preference for affirmative results, reported in Panel~A, Column~3. Under this specification, and assuming no internal bias, the corrected estimate is $-0.068$ with a 95\% confidence interval from $-0.101$ to $-0.036$. We then impose the same four-fold selection ratio but allow affirmative studies to have greater average internal bias than non-affirmative studies -- the mean internal bias is set to $0.05$ for affirmative studies and $0.01$ for non-affirmative studies.\footnote{Here we assume: (1) affirmative studies are four times more likely to be published than non-affirmative studies; and (2) affirmative studies have a mean internal bias of 0.05, and (3) nonaffirmative studies have a mean internal bias of 0.01, which indicates very little bias.} Because the estimates in our application are measured as Cohen's $d$, the internal-bias parameters are also measured in standard-deviation units. Thus, a value of $0.05$ means that affirmative estimates are assumed to contain, on average, internal bias of 0.05 standard deviations, while a value of $0.01$ corresponds to very small average internal bias among non-affirmative estimates. These values should be interpreted as sensitivity parameters that quantify how much the corrected meta-analytic estimate would change under alternative assumptions about study-level bias. 

Under these assumptions, allowing for differential internal bias moves the corrected estimate from the publication-bias-only benchmark of $-0.068$ to $-0.097$, and then to $-0.111$ when the assumed internal bias among affirmative estimates is increased to $0.08$. The multi-bias exercise therefore clarifies that the smaller RE-SR4 estimate reflects the assumption that affirmative and non-affirmative estimates have the same average internal bias. Once affirmative studies are allowed to have somewhat greater internal bias, the corrected estimates move toward the MAIVE and RoBMA estimates rather than toward the smaller RTMA estimate. We therefore interpret the multi-bias results as a sensitivity analysis rather than as a standalone correction. Its main implication is that plausible differences in internal bias across affirmative and non-affirmative estimates do not overturn the conclusion of a substantial negative effect of COVID-19 on learning.

Overall, the methods in this section reinforce the interpretation from Section~\ref{sec:pubbias}. MAIVE yields an estimate almost identical to RoBMA, while RTMA produces a smaller estimate but fits the upper tail of the distribution less accurately. The multi-bias analysis shows that allowing affirmative studies to differ modestly in internal bias moves the selection-adjusted estimates back toward the RoBMA--MAIVE range. We therefore place greatest weight on RoBMA and MAIVE, which point to a learning-loss effect of approximately $-0.12$, or about 30\% of a school year.

\section{Conclusion}
\label{sec:conclusion}

This paper builds on the meta-analysis of \cite{betthauser2023systematic} on the effect of the COVID-19 pandemic on the learning progress of school-aged children, extending their analysis by applying a comprehensive set of formal bias correction methods --- spanning funnel-based tests, selection models, Bayesian meta-analysis, and instrumental variable estimation --- to assess the robustness of the underlying effect size estimate against publication bias and $p$-hacking.

Our graphical analysis reveals that the evidence from visual diagnostics alone is ambiguous. The log-transformed $z$-score distribution shows no clustering around the significance threshold. However, the raw $z$-score distribution exhibits pronounced bunching at the 5\% level, and our funnel plot displays some asymmetry, with a wider spread of estimates on the negative side. Taken together, these patterns are consistent with some degree of selective publication, though not conclusive on their own, and motivate the formal correction methods we employ.

PET-PEESE yields a corrected estimate of $-0.245$, almost double the uncorrected estimate in magnitude, and is consistent with publication bias. The 3PSM and RoBMA estimates, at $-0.123$ and $-0.118$ respectively, are closer to the uncorrected mean of $-0.14$, but still suggest modest downward bias. The sensitivity analyses of \cite{mathur2020sensitivity} show that while results are sensitive to a worst-case selection scenario --- MAN yields a sign reversal to $0.021$ --- an implausibly large degree of bias, 29.60 times stronger publication of affirmative over non-affirmative studies, would be required to drive the point estimate to zero. No degree of selection bias under this model can shift the estimate to $+0.05$.

Accounting additionally for $p$-hacking and the spurious correlation between estimates and standard errors introduced by normalization to Cohen's $d$, MAIVE yields an estimate of $-0.119$, highly statistically significant and closely consistent with RoBMA. The RTMA estimate of $-0.039$, while directionally consistent, is discounted due to imperfect upper-tail fit indicated by the diagnostic Q-Q plot. The multi-bias analysis, which allows for differential internal bias between affirmative and non-affirmative studies, further corroborates the RoBMA and MAIVE estimates: allowing affirmative estimates to carry somewhat larger internal bias moves the corrected estimates toward the RoBMA--MAIVE range, rather than estimating such differential bias from the data.

The key finding is the near-perfect consistency between RoBMA and MAIVE --- both yielding estimates around $-0.119$ --- and their close alignment with the uncorrected benchmark estimate of $-0.14$. RoBMA model-averages across publication-bias models, while MAIVE addresses endogenous precision and the mechanical link between Cohen's $d$ and its standard error. This consistency across methodologically distinct approaches indicates that COVID-19 is associated with a learning deficit of approximately 30\% of a school year based on the preferred RoBMA and MAIVE estimates, compared with about 35\% using the uncorrected benchmark. This magnitude remains economically meaningful: the corrections change the estimated size of the learning deficit, but the shock remains large enough to matter for human-capital policy. While our analysis reveals evidence of selection bias in the underlying data, this bias does not systematically distort the overall findings of \cite{betthauser2023systematic}. The conclusion of a substantial and persistent COVID-19 learning deficit is robust to a wide range of correction methods.  

Bias-correction methods address different forms of bias, rest on different assumptions, and can yield meaningfully different corrected estimates. Credible evidence synthesis therefore requires applying bias-correction methods and assessing how conclusions vary across them. Without such comparisons, it is difficult to know whether a corrected estimate reflects a robust empirical conclusion or primarily the assumptions of the chosen method.

\section*{Data Availability Statement} The effect-size dataset analyzed in this paper was  compiled by \citet{betthauser2023systematic} and is publicly  available at \url{https://doi.org/10.17605/osf.io/u8gaz}.  Replication code for all analyses reported in this  paper is available at \url{https://meta-analysis.cz/learning}.
 
\section*{Use of Generative AI} The authors used generative AI tools, including Claude Sonnet (Anthropic) and ChatGPT (OpenAI), during manuscript preparation to improve language and readability and to assist with formatting and debugging statistical code in \texttt{R}. The tools were accessed through their standard web interfaces during 2025--2026. AI tools were not used for study selection, data coding, estimation choices, analytical decisions, interpretation of results, or the formulation of conclusions. AI assistance with code was limited to formatting, debugging, and implementation support; all statistical analyses were run and verified by the authors. All results reported in the paper are fully reproducible without AI tools. The authors reviewed and verified all AI-assisted output and take full responsibility for the content of the manuscript. AI use follows the guiding principles of \citet{cook2026guidance} and the reporting standards of \citet{cook2026reporting}.

\clearpage

\bibliographystyle{kluwer}
\begingroup
\setstretch{0.5}
\bibliography{biblio} 
\endgroup

\clearpage

\clearpage
\section{Figures} \label{sec:Figures}

\begin{figure}[htbp]
\caption{Effect size estimates by country, ordered by median.} \label{fig:country}
\begin{center}
 \includegraphics[width=0.75\textwidth]{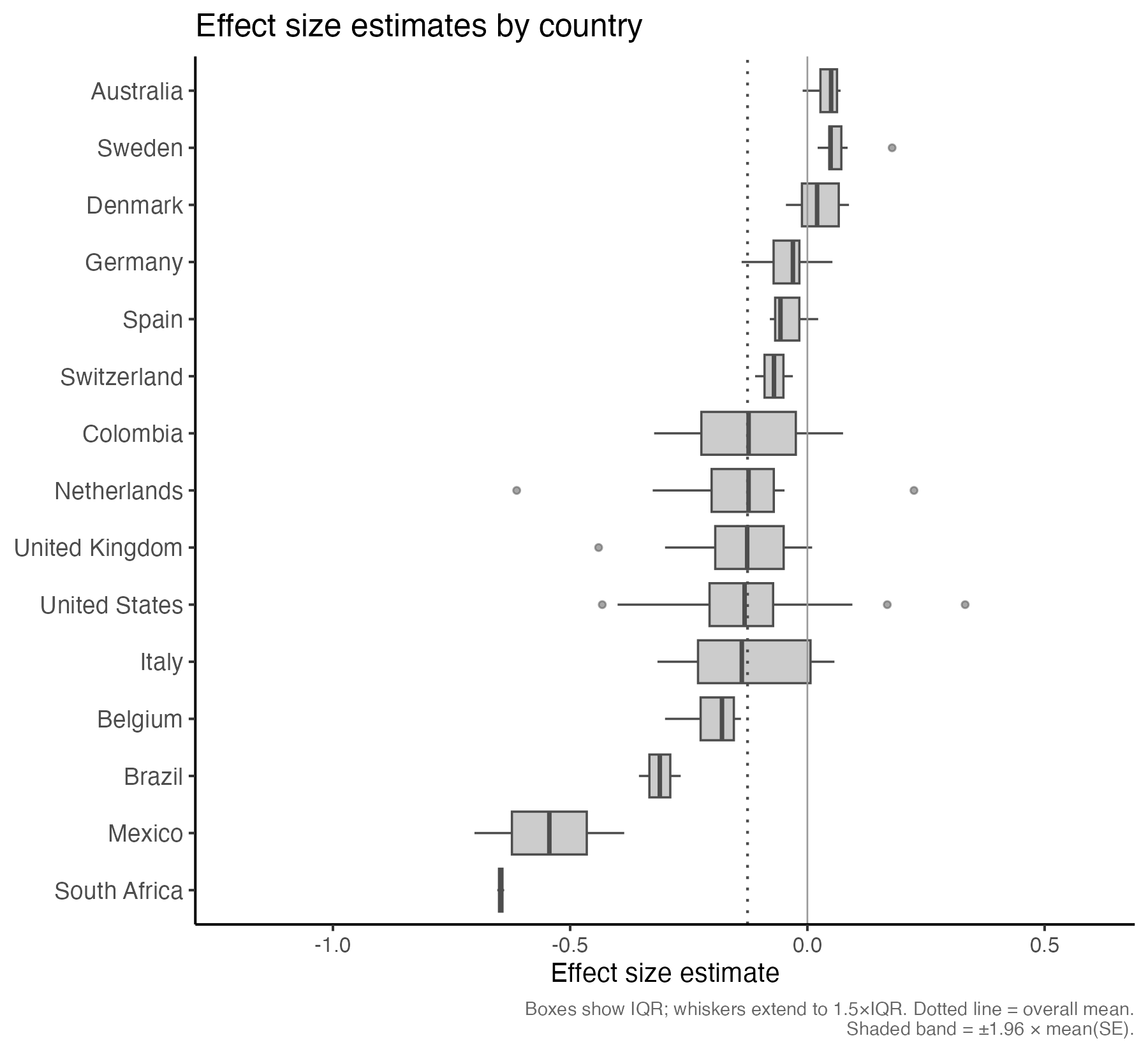}
\end{center} 
\caption*{ 
\textit{Notes:} Boxes show the interquartile range of effect-size estimates within each country; whiskers extend to 1.5 $\times$ IQR, and dots denote outliers. The vertical dotted line marks the overall mean effect. Countries at the top of the figure, including Australia, Sweden, and Denmark, report near-zero or slightly positive effects, while Mexico and South Africa show the largest negative estimates. The three most represented countries---the United States, the United Kingdom, and the Netherlands---cluster close to the overall mean, though with substantial within-country dispersion.
} 
\end{figure}

\begin{figure}
 \caption{Distribution of effect size estimates by subject area and school level.} \label{fig:heterogeneity}
\begin{center}
 \includegraphics[width=12cm]{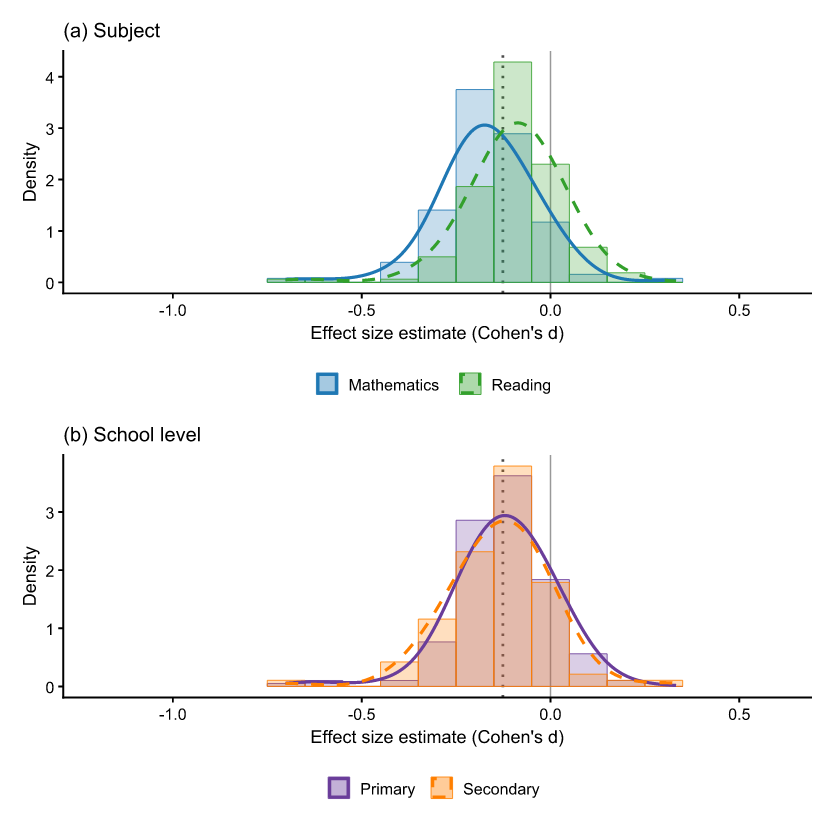}
\end{center}
\caption*{ 
\emph{Note:} Panel (a) shows the distribution of effect sizes for mathematics ($N=128$) and reading ($N=161$), and panel (b) for primary ($N=196$) and secondary education ($N=95$). Each panel overlays histograms and kernel density curves. The solid vertical line marks zero (no deficit); the dotted vertical line marks the overall pooled mean ($d = -0.126$). Learning deficits in mathematics tend to be larger on average than in reading (unweighted means of $-0.167$ vs. $-0.095$), whereas primary and secondary education show similar distributions ($-0.122$ vs. $-0.135$). Two estimates covering a composite of mathematics and reading are excluded from panel (a).
}
\end{figure}

\begin{figure}[htbp]
\caption{Effect size estimates by grade, with a loess smoother and 95\% confidence band.} \label{fig:grade}
\begin{center}
 \includegraphics[width=\textwidth]{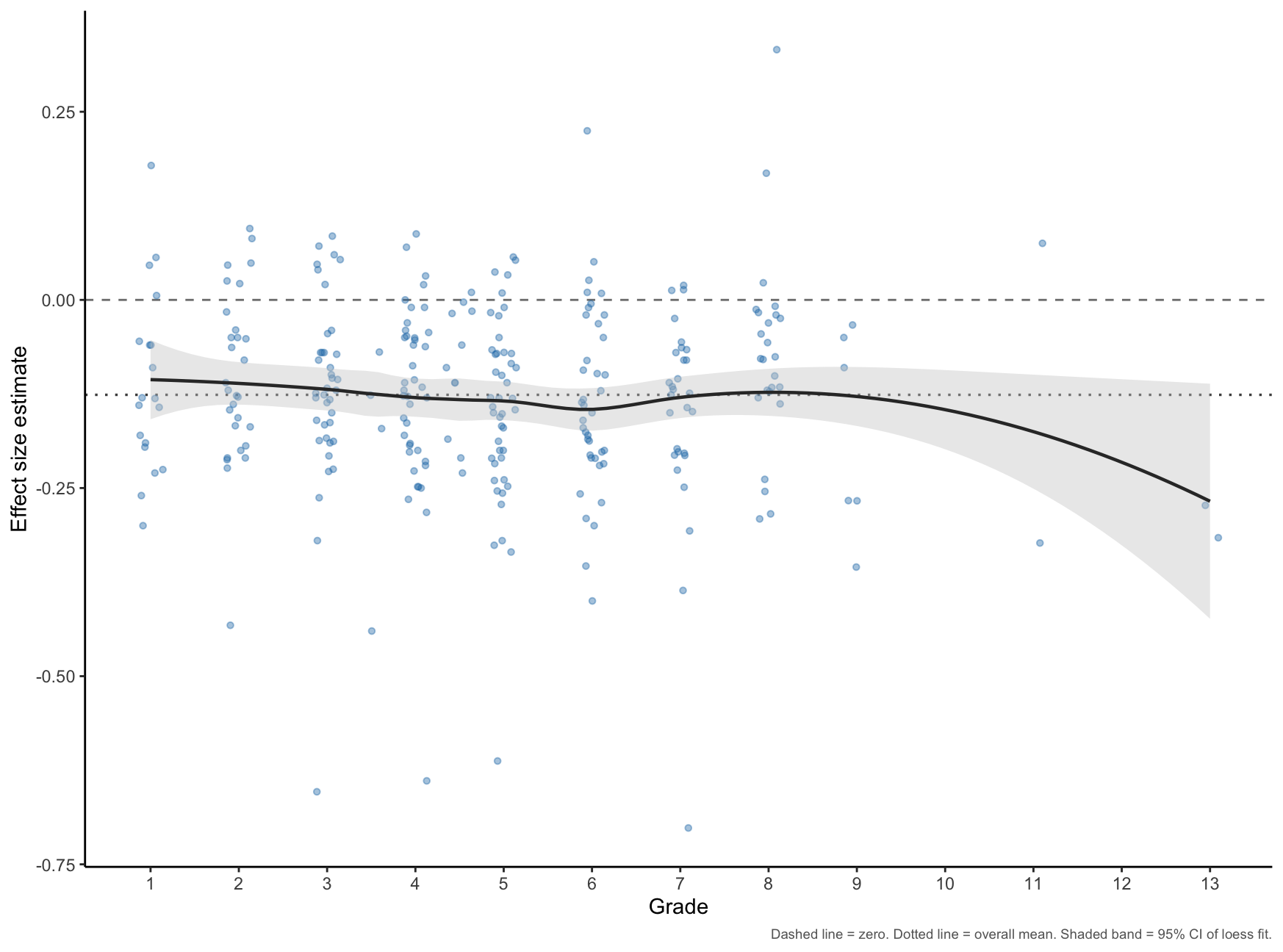}
\end{center}
\caption*{ 
\emph{Note:} Individual estimates are shown as jittered points. The dashed line marks zero; the dotted line marks the overall mean ($-0.126$). Effects are broadly stable across grades 1--9, consistent with the absence of a meaningful school-level gradient visible in panel~(b) of Figure~\ref{fig:heterogeneity}. The apparent decline in grades 11--13 should be interpreted cautiously: these grades are represented by very few estimates (two at grade 11, two at grade 13), and the wide confidence band reflects this uncertainty.
}
\end{figure}

\begin{figure}[htbp]
 \caption{\small Conventional funnel plot} \label{fig:funnel} 
 \begin{center}
 \includegraphics[width=0.7\linewidth, clip]{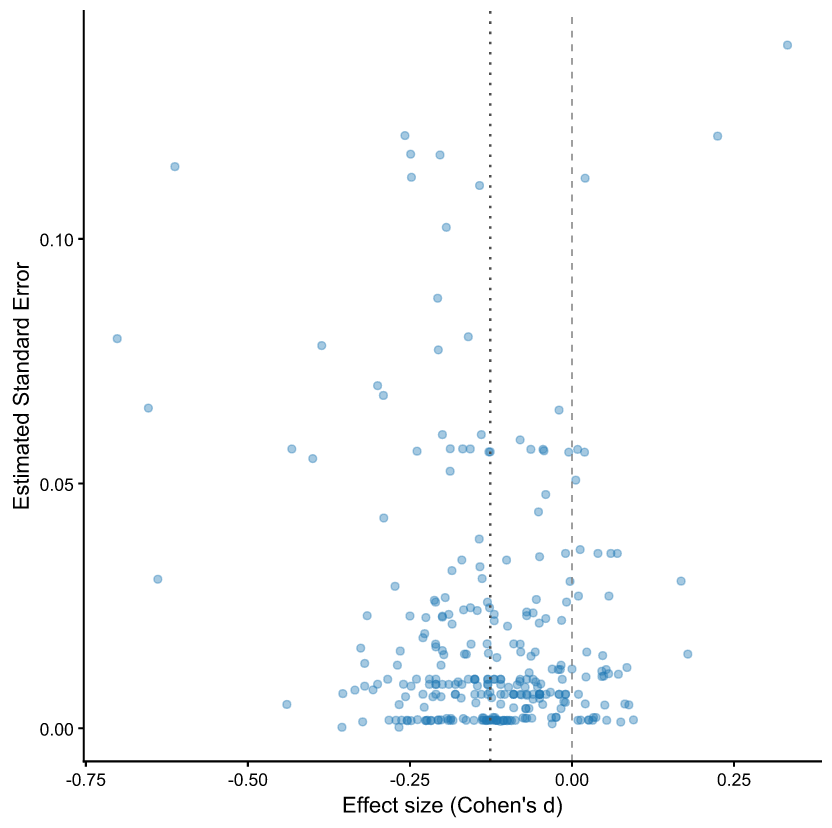}
 \end{center}
\caption*{ 
\emph{Notes:} The figure shows a conventional funnel plot with effect sizes on the horizontal axis and their estimated standard errors on the vertical axis. The dotted vertical line marks the pooled mean effect. The dashed vertical line marks zero.
}

\end{figure}

\begin{figure}[htbp]
\caption{\small Distribution of effect size estimates by statistical significance} \label{fig:significantvsnot}
 \begin{center}
 \includegraphics[width=0.8\linewidth, trim={0 0 0 15mm}, clip]{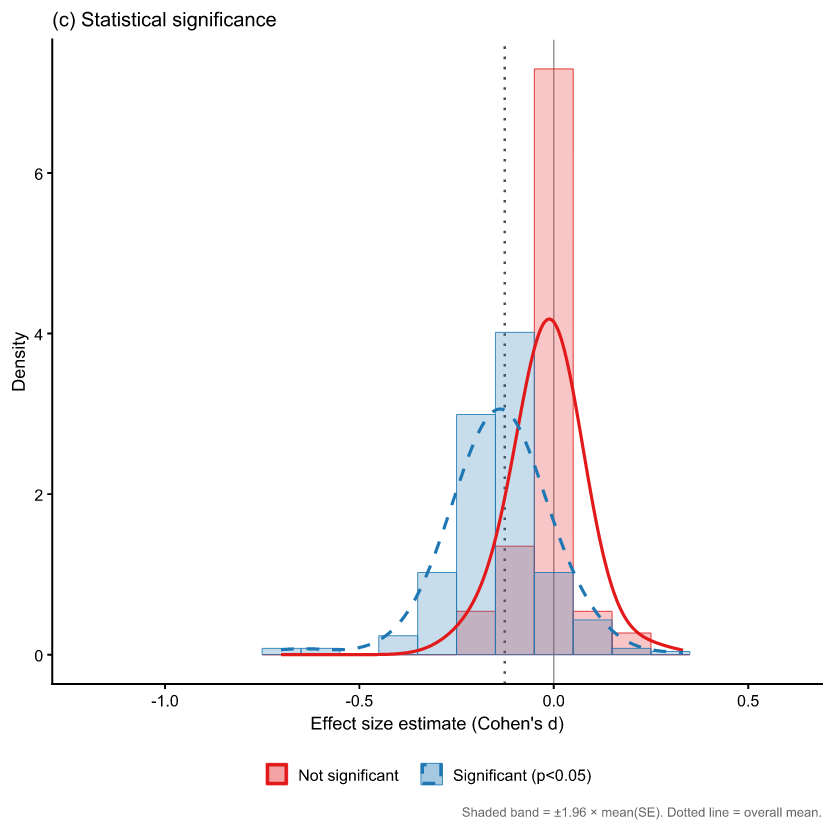}
 \end{center}
\caption*{ 
\emph{Notes:} The figure shows the distribution of effect size estimates (Cohen’s $d$), separately for statistically significant ($p<0.05$, blue) and non-significant estimates (red). The dotted vertical line marks the overall mean effect size. The shaded band indicates the interval defined by $\pm 1.96 \times$ the mean standard error around zero.
} 
\end{figure} 

\begin{figure}[htbp]
\vspace{-1cm}
 \caption{Significance funnel plot}\label{fig:sig_funnel}
 \begin{center}
 \includegraphics[width=0.7\textwidth]{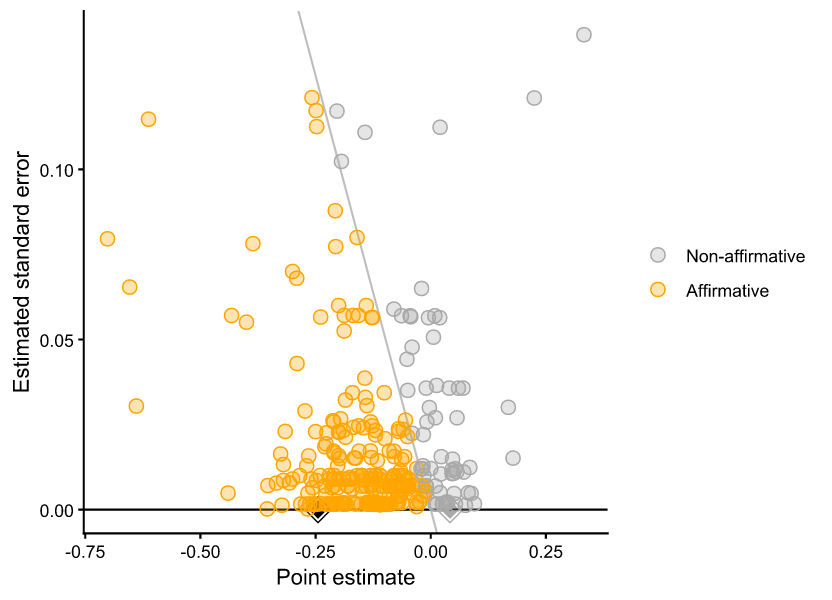}
 \hfill 
 \end{center} 
 \vspace{-0.5cm}
\caption*{  
 \emph{Notes:} The figure displays a significance funnel plot following  \citet{mathur2020sensitivity}. Orange points are affirmative estimates;  gray points are non-affirmative. The gray diamond is the fixed-effects  mean among non-affirmative estimates ($0.021$), corresponding to the MAN  worst-case benchmark reported in Panel~A of Table~\ref{tab:phack}. The  black diamond is the precision-weighted pooled mean across all 291 estimates  ($\hat{\mu}_{IVW} = -0.245$); it lies more negative than the Panel~A  equal-weighted mean of $-0.126$ because inverse-variance weighting gives  disproportionate influence to large, high-precision studies.
}
\end{figure}

\begin{figure}[htbp]
\vspace{-0.5cm}
 \caption{Distribution of $z$-scores.}\label{fig:diagnostic_z}
 \begin{center}
 \includegraphics[width=0.7\textwidth, trim= 0 0  0 1cm, clip]{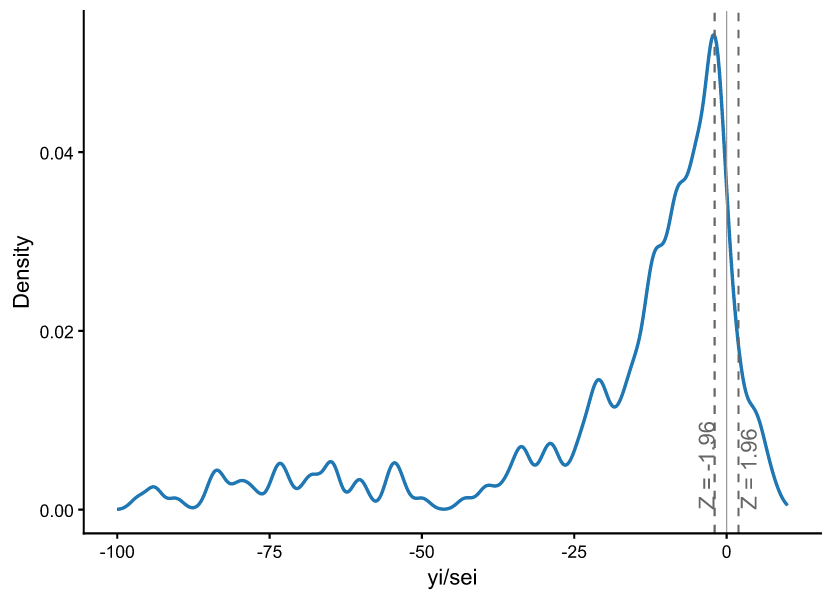} 
 \end{center}
 \vspace{-0.5cm}        
\caption*{ \emph{Note:} The figure shows the distribution of $z$-scores, $y_i/se_i$, in the original dataset. The dashed vertical lines mark $z=-1.96$ and $z=+1.96$, the conventional 5\% two-sided significance thresholds. 
}
\end{figure}

\begin{figure}[htbp] 
 \caption{Comparison between the distribution of the $z$-scores in absolute values (a), (b), and in logs of absolute values (c).}\label{fig:comparison} 
 \vspace{-0.5cm} 
 \begin{center}
 \subfigure[The distribution of absolute $z$-scores, $abs(y_i/se_i)$. The dashed vertical line marks $z = 1.96$, the conventional 5\% two-sided significance threshold.]{\includegraphics[width=0.75\textwidth, trim= 0 1.5cm  0 1.5cm, clip]{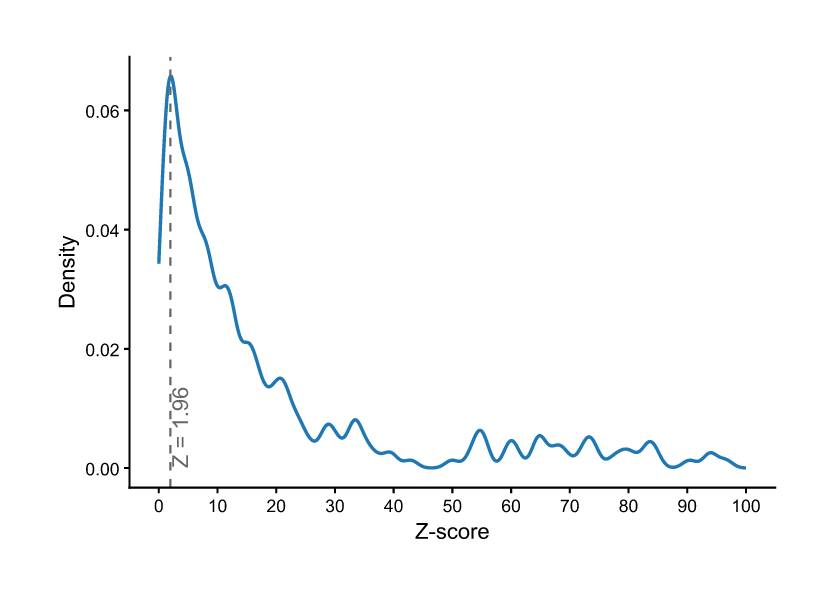}} 
  \subfigure[The distribution of absolute $z$-scores, ($abs(y_i/se_i)$), zoomed to the (0, 20) neighborhood. The dashed vertical line marks $z = 1.96$, the conventional 5\% two-sided significance threshold.]
 {\includegraphics[width=0.75\textwidth, trim= 0 2cm  0 2cm, clip]{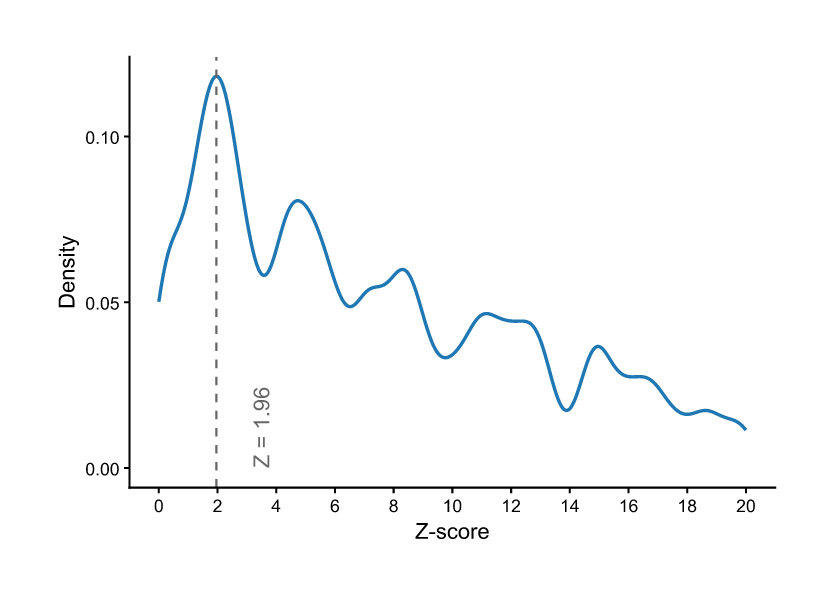}}
 \end{center} 
\end{figure}

\begin{figure}[htbp]
\ContinuedFloat
 \caption{(continued)}\label{fig:comparisonzoomed}
 \vspace{-0.5cm} 
 \begin{center}
 \subfigure[The figure shows the distribution of log absolute $z$-scores, $ln(abs(y_i/se_i))$. The dashed vertical line marks $z = 1.96$, the conventional 5\% two-sided significance threshold.]{\includegraphics[width=0.7\textwidth, trim= 0 2cm  0 2cm, clip]{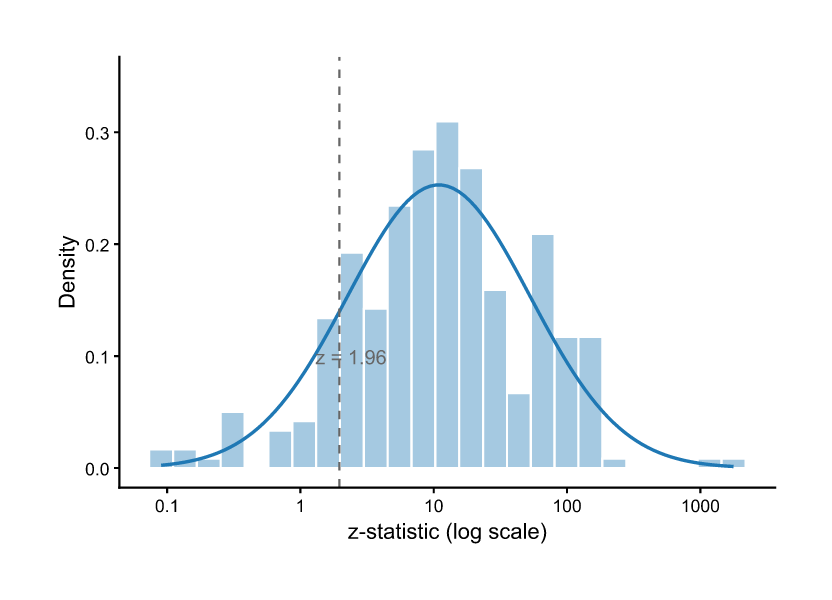}} 
 \end{center} 
\end{figure}

\begin{figure}[htbp] 
    \caption{DFBETAS influence diagnostics for intercept. }
    \label{fig:dfbetas}
    \begin{center} 
            \includegraphics[width=\textwidth]{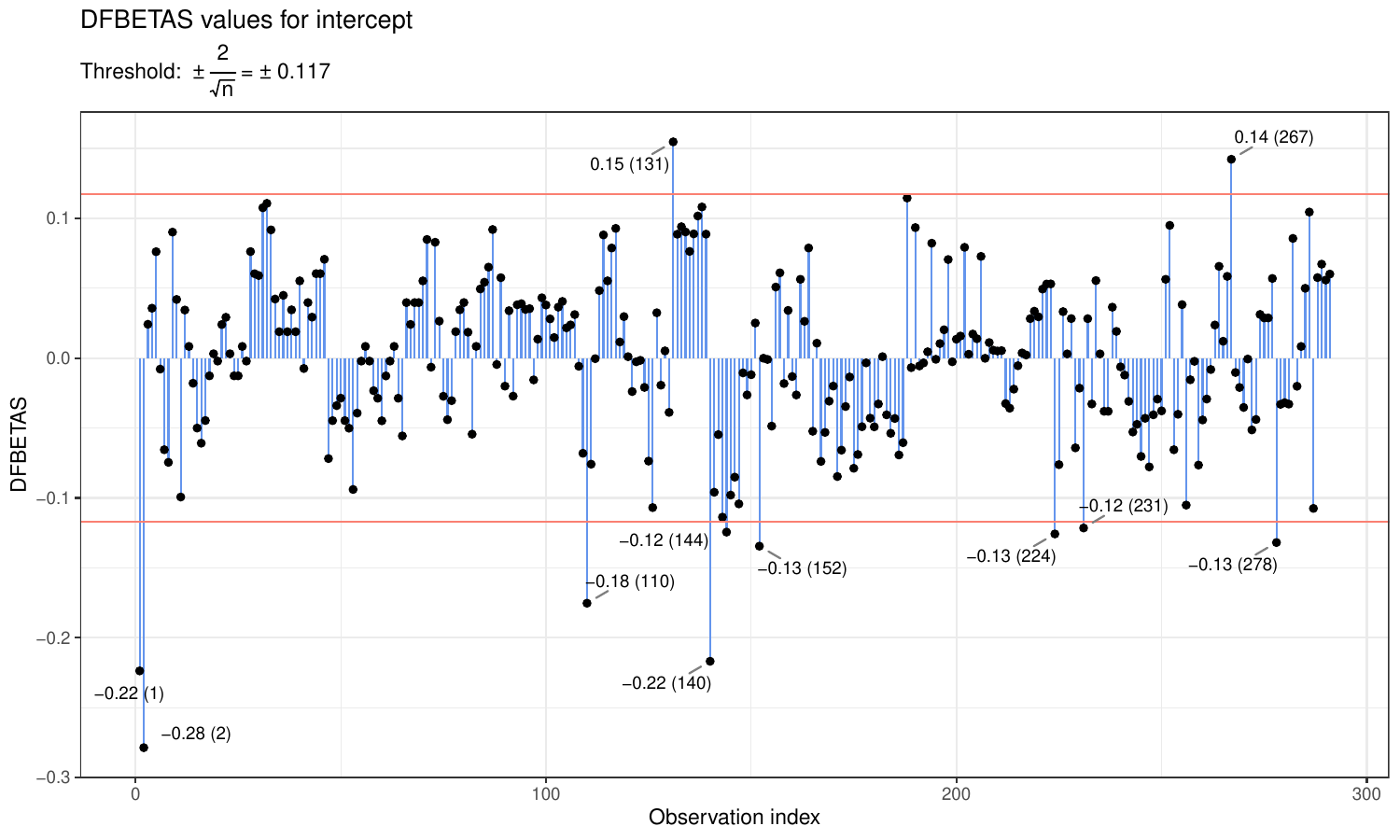}
    \end{center}
    \begin{footnotesize}
       \emph{Notes:} The figure displays DFBETAS values for all 291 estimates,  measuring the change in the pooled random-effects estimate when each observation is omitted in turn. The horizontal lines mark the sample-size-adjusted threshold $\pm 2/\sqrt{n} = \pm 0.117$. Eleven estimates exceed the threshold; excluding them moves the pooled estimate from $-0.124$ to $-0.116$. 
    \end{footnotesize}
\end{figure}

\begin{figure}[htbp]
 \caption{Diagnostic $q$-$q$ plot 
 } \label{fig:diagnostic_qq}
 \begin{center}
 \includegraphics[width=0.63\textwidth, trim={0 0 0 10mm}, clip]{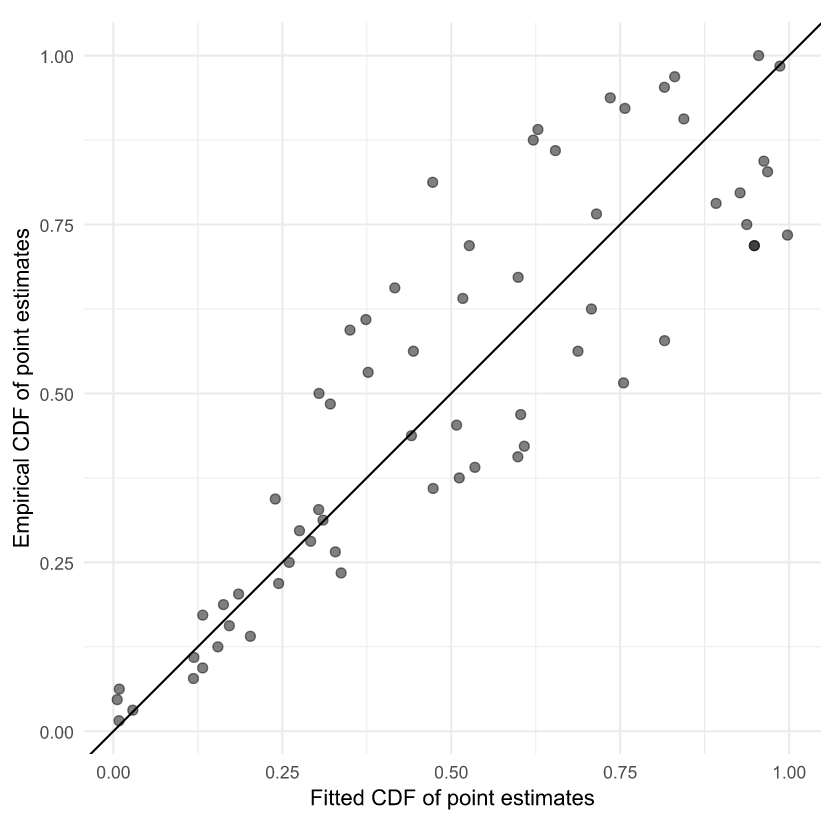}
 \hfill 
 \end{center}  
\caption*{ \emph{Notes:}  
 The figure displays the RTMA diagnostic Q-Q plot, comparing the fitted CDF of published non-affirmative estimates (horizontal axis) with their empirical CDF (vertical axis). Points lying on the 45-degree line indicate good model fit. Deviations in the upper tail suggest the truncated normal model fits the lower quantiles well but captures the upper tail less accurately, consistent with the mechanical distortion introduced by standardizing estimates as Cohen's $d$. } 
\end{figure}


\clearpage
\section{Tables}
\label{sec:Tables}
\begin{footnotesize}
    
\vspace{-0.2cm}
     
\begin{table}[htbp] 
\centering
\footnotesize 
\begin{threeparttable}

\caption{Descriptive statistics}
\label{tab:dscrstat}
\begin{tabularx}{\textwidth}{@{}>{\raggedright\arraybackslash}Xcccccc@{}} 
\toprule
Variable & $N$ & Mean & SD & Min & Median & Max \\
\hline
Standardized mean difference & 291 & $-0.126$ & $0.127$ & $-0.702$ & $-0.124$ & $0.333$ \\
Standard error & 291 & $0.019$ & $0.026$ & $< 0.001$ & $0.009$ & $0.140$ \\
Sample size ($n$) & 291 & 446{,}417 & --- & 275 & 50{,}000 & 10{,}884{,}922 \\
Absolute $z$-statistic & 291 & $37.8$ & $133.7$ & $0.000$ & $11.0$ & $1{,}775$ \\ 
\hline
Composition &&&&&& \\
\hline
Primary education & 196 & \multicolumn{5}{l}{share = 67\%} \\
Secondary education & 95 & \multicolumn{5}{l}{share = 33\%} \\
Mathematics & 128 & \multicolumn{5}{l}{share = 44\%} \\
Reading & 161 & \multicolumn{5}{l}{share = 55\%} \\
Mixed & 2 & \multicolumn{5}{l}{share = 0.69\%} \\
\bottomrule
\end{tabularx}
\begin{tablenotes}[flushleft]
\footnotesize
 \item[] \emph{Notes:} The dataset contains 291 effect size estimates from 42 studies across 15 countries, drawn from \citet{betthauser2023systematic}. Effect sizes are expressed as standardized mean differences. Sample size refers to the number of students in the primary study from which each estimate is drawn. The mean and median sample size reflect the highly right-skewed distribution of study sizes; two estimates covering both mathematics and reading jointly are excluded from the subject composition rows.
\end{tablenotes}

\end{threeparttable}

\end{table}

\vspace{-.2cm}
    \begin{table}[htbp]
\centering
\caption{Summary statistics by subgroup}\label{tab:subgroups}
\centering
\resizebox{\ifdim\width>\linewidth\linewidth\else\width\fi}{!}{
\begin{tabular}[t]{llllllll}
\toprule
\multicolumn{2}{c}{ } & \multicolumn{3}{c}{Unweighted} & \multicolumn{3}{c}{Weighted (1/n per study)} \\
\cmidrule(l{3pt}r{3pt}){3-5} \cmidrule(l{3pt}r{3pt}){6-8}
& N & Mean & \multicolumn{2}{c}{95\% CI} & Mean & \multicolumn{2}{c}{95\% CI}  \\
\midrule
\textit{Subject} &  &  &  &  &  &  & \\
\hspace{1em}Mathematics & 128 & -0.167 & -0.188 & -0.145 & -0.164 & -0.205 & -0.122\\
\hspace{1em}Reading & 161 & -0.095 & -0.113 & -0.076 & -0.13 & -0.169 & -0.09\\
\hspace{1em}Other/mixed & 2 & -0.07 & -0.148 & 0.008 & -0.07 & -0.126 & -0.015\\
\addlinespace
\textit{School level} &  &  &  &  &  &  & \\
\hspace{1em}Primary & 196 & -0.122 & -0.139 & -0.104 & -0.138 & -0.168 & -0.107\\
\hspace{1em}Secondary & 95 & -0.135 & -0.162 & -0.109 & -0.153 & -0.213 & -0.094\\
\addlinespace
\textit{Grade group} &  &  &  &  &  &  & \\
\hspace{1em}Early primary (1--4) & 132 & -0.119 & -0.14 & -0.098 & -0.144 & -0.186 & -0.102\\
\hspace{1em}Middle (5--8) & 149 & -0.128 & -0.149 & -0.108 & -0.132 & -0.17 & -0.093\\
\hspace{1em}Secondary (9+) & 10 & -0.19 & -0.283 & -0.097 & -0.199 & -0.312 & -0.086\\
\addlinespace
\textit{Sample size} & & & & & & & \\
\hspace{1em}Small ($<$10k) & 74 & -0.151 & -0.192 & -0.111 & -0.202 & -0.267 & -0.138\\
\hspace{1em}Medium (10k--100k) & 118 & -0.108 & -0.126 & -0.091 & -0.094 & -0.115 & -0.074\\
\hspace{1em}Large ($>$100k) & 99 & -0.129 & -0.15 & -0.107 & -0.131 & -0.17 & -0.092\\
\addlinespace
\textit{Country} &  &  &  &  &  &  & \\
\hspace{1em}United States & 149 & -0.136 & -0.153 & -0.12 & -0.142 & -0.166 & -0.117\\
\hspace{1em}United Kingdom & 58 & -0.128 & -0.151 & -0.104 & -0.136 & -0.165 & -0.107\\
\hspace{1em}Netherlands & 27 & -0.146 & -0.198 & -0.093 & -0.14 & -0.187 & -0.093\\
\hspace{1em}Other countries & 57 & -0.089 & -0.138 & -0.039 & -0.147 & -0.207 & -0.086\\
\addlinespace
\textit{Source tier} &  &  &  &  &  &  & \\
\hspace{1em}Tier 1: Peer-reviewed & 62 & -0.11 & -0.157 & -0.062 & -0.165 & -0.233 & -0.098\\
\hspace{1em}Tier 2: Working paper & 65 & -0.132 & -0.164 & -0.1 & -0.13 & -0.164 & -0.096\\
\hspace{1em}Tier 3: Gov't / institutional & 34 & -0.12 & -0.156 & -0.084 & -0.122 & -0.165 & -0.079\\
\hspace{1em}Tier 4: Commercial & 130 & -0.133 & -0.148 & -0.118 & -0.138 & -0.163 & -0.113\\
\addlinespace
\textbf{All estimates} & 291 & -0.126 & -0.141 & -0.112 & -0.142 & -0.17 & -0.115\\

\bottomrule
\end{tabular}}
\end{table}

    \begin{table}[htbp]
\begin{center}
 \caption[Publication bias]{Publication bias: baseline and corrected estimates}
 \label{tab:pubbias}
\footnotesize
\begin{threeparttable} 
\begin{tabular}{@{}lcccc@{}}
\toprule \\

\multicolumn{5}{l}{Panel A: Baseline mean estimates}\\
\midrule
 & \textbf{Unweighted} & \textbf{RVE} & &\\
Mean effect & $-0.126$** & $-0.140$*** & &\\
 & ($0.059$) & ($0.020$) & &\\[0.3cm]
 \midrule
Observations & $291$ & $291$ & &\\ 
\bottomrule \\ 

\multicolumn{5}{l}{Panel B: Corrected estimates}\\
\midrule
 & \textbf{PET} & \textbf{PEESE} & \textbf{3PSM} & \textbf{RoBMA}\\
Effect beyond bias & $-0.271$*** & $-0.245$*** & $-0.123$*** & $-0.118$\\
 & ($0.040$) & ($0.051$) & ($0.009$) 
 & [$-0.135$, $-0.094$]\\[0.3cm]
Publication bias & $29.269$*** & & &\\
 & ($9.818$) & & &\\[0.3cm]
Publication bias & & $114.145$ & &\\
\textit{Standard error$^2$} & & ($58.836$) & &\\[0.3cm]
Likelihood ratio test & & & $\chi^2 = 0.034$ &\\
\textit{H0: no pub. bias} & & & $p$-value $= 0.854$ &\\[0.3cm]
\midrule
Observations & $291$ & $291$ & $291$ & $291$\\
\bottomrule
 
\end{tabular}
\begin{tablenotes}[flushleft]
 \footnotesize
 
 \item[] \emph{Notes: } Unweighted = the equal-weighted (arithmetic) mean of all 291 estimates, i.e.\ the simple estimate-level mean, not the inverse-variance fixed-effect estimator. RVE = robust variance estimation mean following \cite{hedges2010robust}, clustering at the study level to account for within-study dependence; the RVE estimate of $-0.140$ coincides with the pooled mean reported by \cite{betthauser2023systematic}. PET = precision effect test based on the estimates of weighted regression $estimate_{ij}=\beta_0+\beta_1*(SE_{estimate})_{ij}+u_{ij}$, where $estimate_{ij}$ is the $i$-th estimate from the $j$-th study, with $(SE_{estimate})_{ij}$ the respective standard error. PEESE = precision effect estimate with standard errors; for PEESE $(SE_{estimate})_{ij}$ is squared. 3PSM is a publication selection model as in \cite{iyengar1988selection, hedges1992selectionmodeling, vevea1995general}. RoBMA = robust Bayesian model averaging as described in \cite{bartovs2023robustBMA, maier2023robustBMA}. For PET \& PEESE, we report heteroskedasticity robust standard errors clustered at the study level. Significant at $^{***}[1\%]$, $^{**}[5\%]$, $^{*}[10\%]$ level.
 \end{tablenotes}
\end{threeparttable} 
\end{center}
\end{table}

    \begin{table}[htbp]
\begin{center}
    \caption{Correcting for publication bias, additional specifications}
    
\footnotesize
\begin{threeparttable} 
\begin{tabular}{@{}lcccc@{}}
\toprule \\ 
\multicolumn{5}{l}{Panel A: Worst-case and selection-ratio-adjusted estimates}\\
\midrule
                    &  \textbf{MAN} &\textbf{FE-SR4}    & \textbf{RE-SR4}    &   \\
Effect beyond bias  & $0.021$*  & $-0.206$***         & $-0.068$***                &     \\
                    & ($0.012$)  & ($<0.001$)          &  ($0.015$)              &    \\ 
                                        & & & &\\ 
Observations        &$41$ &$291$&$291$ &    \\ 

\midrule
\multicolumn{5}{c}{}\\
\multicolumn{5}{l}{Panel B: Publication bias required to explain away the results}\\
\midrule
& \textbf{\textit{coef}=0} & \textbf{\textit{coef}=0.01}& \textbf{\textit{coef}=0.05} &\\
Estimate's $s$-value &  $29.60$ & $60.65$ &no amount  &  \\
CI $s$-value&  $8.31$ &  $10.46$ & no amount & \\ 
                    & & & &\\ 
Observations & $291$ &$291$& $291$& \\ 

\midrule
\multicolumn{5}{c}{}\\
\multicolumn{5}{l}{Panel C:  $p$-hacking \& multibias}\\
\midrule
                    &\textbf{MAIVE} &  \textbf{RTMA}          &$\mathbf{Multi}_{0.05;0.01}$ & $\mathbf{Multi}_{0.08;0.01}$     \\
Effect beyond bias  &  $-0.119$*** & $-0.039$***                &$-0.097$***&  $-0.111$***   \\
                    &  ($0.012$) & ($0.003$)                    &  ($0.008$)   &   ($0.008$)    \\ 
  & $[-0.137, -0.087]$  &     &    & \\
    & &&&\\ 
Coefficient on fitted $SE^2$   &     $-0.245$*** &    &         & \\ 
                     & ($0.051$) &     &    & \\ 
               
                     & &&&\\ 
First stage F-statistic & $271.716$    &     &    & \\                
                    & &&&\\ 
Heterogeneity       &  & $0.072$*** &&\\
                    & & ($0.001$) &&\\ 
                    & & &&\\ 
Observations        & $291$ &$291$ &$291$&$291$\\  

\bottomrule 
\end{tabular}
\begin{tablenotes}[flushleft]
      \footnotesize
      \item[] \emph{Notes: } FE = fixed effects mean, RE = mean effect estimated using robust random effects accounting for heterogeneity and clustering, MAN = meta-analysis of non-affirmative studies, FE-SR4 = fixed-effects meta-analysis with a 4-fold preference (selection ratio = 4) for affirmative studies, RE-SR4 = robust random-effects specification accounting for heterogeneity and clustering with the 4-fold preference for affirmative studies, RTMA = right-truncated meta-analysis, MAIVE = meta-analysis instrumental variable estimator, Anderson-Rubin 95\% confidence interval is reported in square brackets, Multi$_{0.05;0.01}$ = affirmative studies bias is set to 0.05 and non-affirmative studies bias to 0.01, Multi$_{0.08;0.01}$ = affirmative studies bias is set to 0.08 and non-affirmative studies bias to 0.01. Standard errors are reported in parentheses. Significant at $^{***}[1\%]$, $^{**}[5\%]$, $^{*}[10\%]$ level. \cite{mathur2020sensitivity, mathur2024p, mathur2024internal, irsova2025MAIVE}
    \end{tablenotes} 
\end{threeparttable} 
\label{tab:phack}
\end{center}
\end{table}

\end{footnotesize}

\clearpage

\renewcommand{\thefigure}{A\arabic{figure}}
\renewcommand{\thetable}{A\arabic{table}}
\setcounter{figure}{0}
\setcounter{table}{0}
\clearpage 
\appendix  
\setcounter{page}{1}   


\clearpage
\begingroup
\thispagestyle{empty}

\begin{center}
\vspace*{2cm}

{\Large\textbf{Online Appendix to:}}\\[0.8em]

{\Large\textbf{Publication bias and $p$-hacking in the effect of COVID-19 on learning}}\\[2em]

Martina Luskova, Nino Buliskeria, Ali Elminejad, Tomas Havranek,\\
Zuzana Irsova, Stepan Jurajda, Marek Kapicka

\end{center}

\endgroup

\section{Computational Reproducibility of \texorpdfstring{\citet{betthauser2023systematic}}{Betthäuser et al. (2023)}}

\label{app:reproduction}

In this section, we report on the computational reproducibility of the meta-analysis by \citet{betthauser2023systematic}. Because our analysis relies on their study sample and dataset, rather than on an independent literature search, we do not provide a separate PRISMA flow diagram. Readers are referred to their Figure 1, which documents the study identification and selection process following PRISMA guidelines. \autoref{tab:reppkg_table} summarizes the contents of the replication package. 
\subsection{Computational Reproducibility}

We used the replication package available on the \href{https://doi.org/10.17605/osf.io/u8gaz}{Open Science Framework}.\footnote{\url{https://doi.org/10.17605/osf.io/u8gaz}} The replication package contains both code and data. The code incorporates the cleaning of the provided data. The final analysis data can be downloaded directly using the code in the replication package. See \autoref{tab:reppkg_table} for the description of replication package contents and reproducibility. We successfully computationally reproduced all the main results (\textit{i.e.}, Figures 2b (pg. 377), 3 (pg. 378), 4 (pg. 379), and 6 (pg. 380)) from the raw data. The remaining figures were compiled manually by the original authors. Originally, Figure 4, pg. 379, is generated using an R extension in STATA. We reproduced the figure using the same code directly in R. In \autoref{sec: app_figures}, we present the reproduced figures; tables are presented in \autoref{sec: app_tables}. \autoref{tab:timeslope} shows the original and replicated slope coefficient estimate, $p$-value, and $95\%$ confidence interval for the learning deficits in time (mentioned in Figure 4, pg.379). \autoref{tab:moderators} shows the original and the reproduced variation in estimates of learning deficit by school subject, level of education, and country income level (original results described in Figure 6, mean differences in text, pg. 380).

\subsection{Discrepancies Between Pre-analysis Plan and Article} 

The authors registered a pre-analysis plan in the \href{https://www.crd.york.ac.uk/prospero/display_record.php?ID=CRD42021249944}{PROSPERO registry}.\footnote{\url{https://www.crd.york.ac.uk/prospero/display_record.php?ID=CRD42021249944}} The paper follows the strategy specified in the pre-analysis plan, and relies on the pre-specified academic and pre-print databases. Regarding the data extraction, we find a minor deviation from the described plan. Despite the pre-analysis plan aiming to collect the key characteristics of the studied countries, the final data set only codes the country names. Moreover, the authors aimed to include the countries' income levels based on the World Bank's classification (low, lower-middle, upper-middle, and high-income). However, the majority of the dataset falls into the high-income category. The upper-middle-income is represented by less than $3\%$ of the data. Low and lower-middle-income categories are not represented at all. Similarly, the final dataset does not include data on the funding source, sample restrictions, survey attrition, and follow-up period(s).

The pre-analysis plan describes the data synthesis strategy but does not specify the standardization measure (the article uses Cohen's \textit{d}). The pre-analysis plan additionally aims to evaluate the learning differences between genders and varying exposure to school closures. These subgroup analyses were not performed due to the unavailability of the data. Lastly, there is no description of the specific tests the authors aimed to conduct. The article includes the following tests. To test for publication bias, the authors use a graphical test based on the distribution of $z$-statistics (assuming that the presence of publication bias can be seen in a notable jump in the distribution of $z$-statistics at the significance threshold, $z = 1.96$ or $p$-value $= 0.05$). Additionally, the supplementary material features two more visual tests: a funnel plot and a test based on the $p$-curve. The article estimates the overall pooled effect size, focuses on the effect size in time, and performs sub-group analysis concerned with socio-economic inequality, school subjects (mathematics and reading), level of education, and country income level.

\clearpage

\subsection{Figures}\label{sec: app_figures}  
  \begin{center}
    \rotatebox{90}{%
    \begin{minipage}{0.85\textheight} 
        \centering
        \captionof{figure}{Publication bias: distribution of $z$-scores}
        \label{fig:z-scors-orig}
        \subfigure[Original Figure 2b]{%
            \includegraphics[trim={1.5cm 0cm 0cm 0cm}, clip, width=0.45\textwidth]{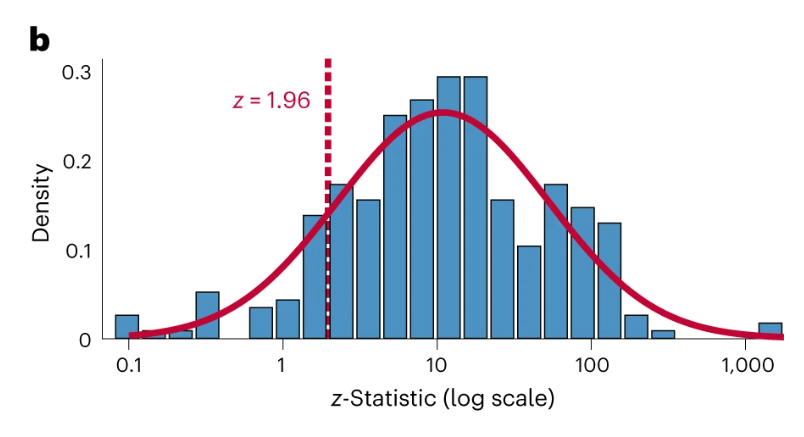}}
        \subfigure[Reproduction of Figure 2b]{%
            \includegraphics[trim={.53cm 0cm 0cm 0cm}, clip, width=0.45\textwidth]{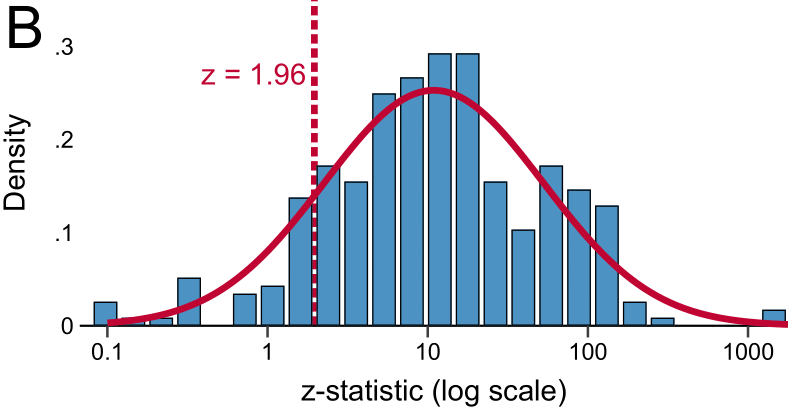}}
        \vspace{1em}
        \captionof*{figure}{{\textbf{Notes:}} The figure displays a visual test for publication bias based on the distribution of the $z$-scores. (a) shows the original, and (b) is the reproduction. There is no difference between the two.}
    \end{minipage}
    } 
\end{center}

\newgeometry{top=1in, bottom=1in, left=.5in, right=.5in}
\begin{figure}[p]
    \caption{Forest plot}\label{fig:forest_replication}
    \hfill
    \subfigure[Original Figure 3]{\includegraphics[width=0.49\textwidth]{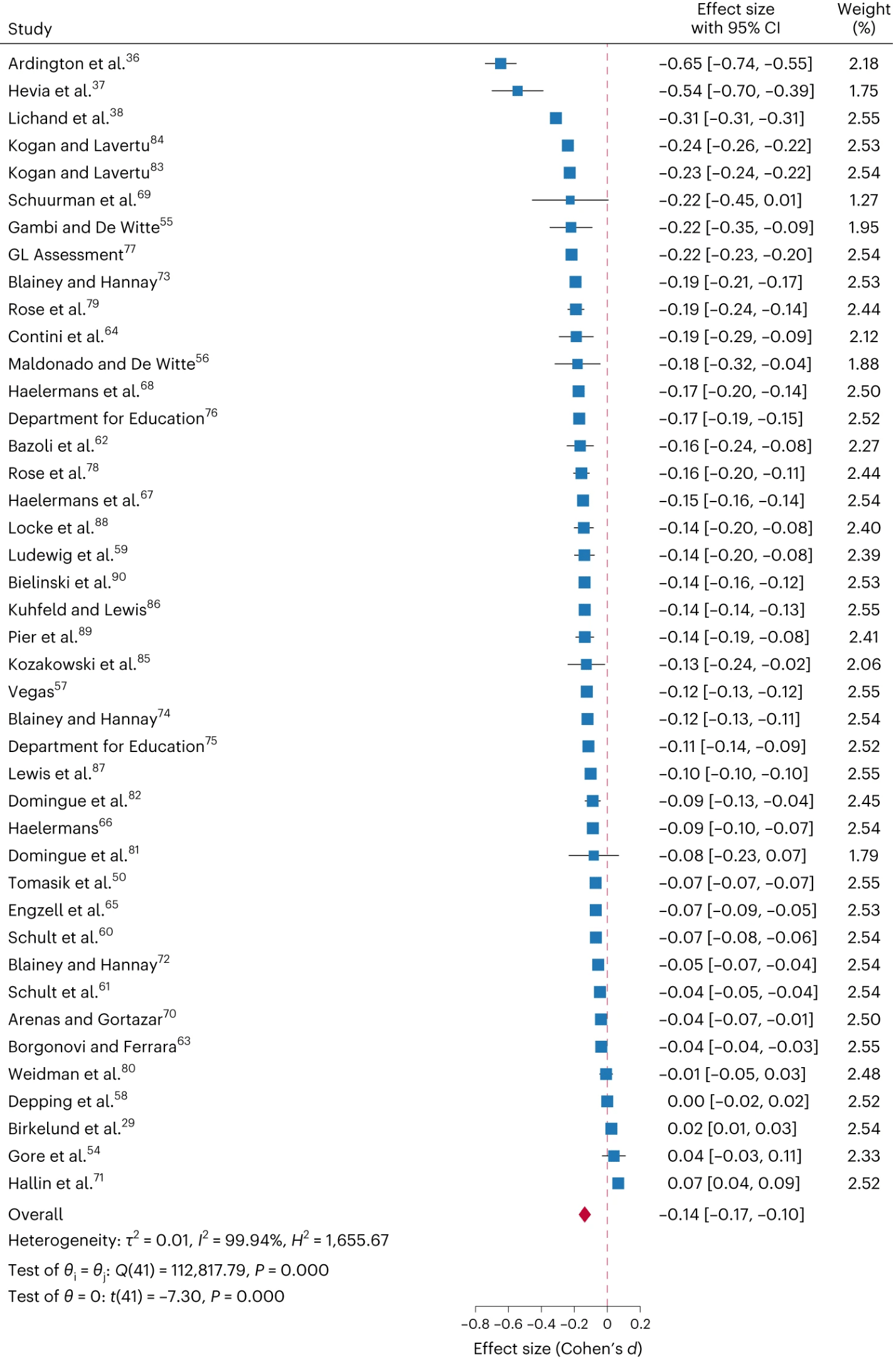}}
    \hfill
    \subfigure[Reproduction of Figure 3]{\includegraphics[width=0.49\textwidth]{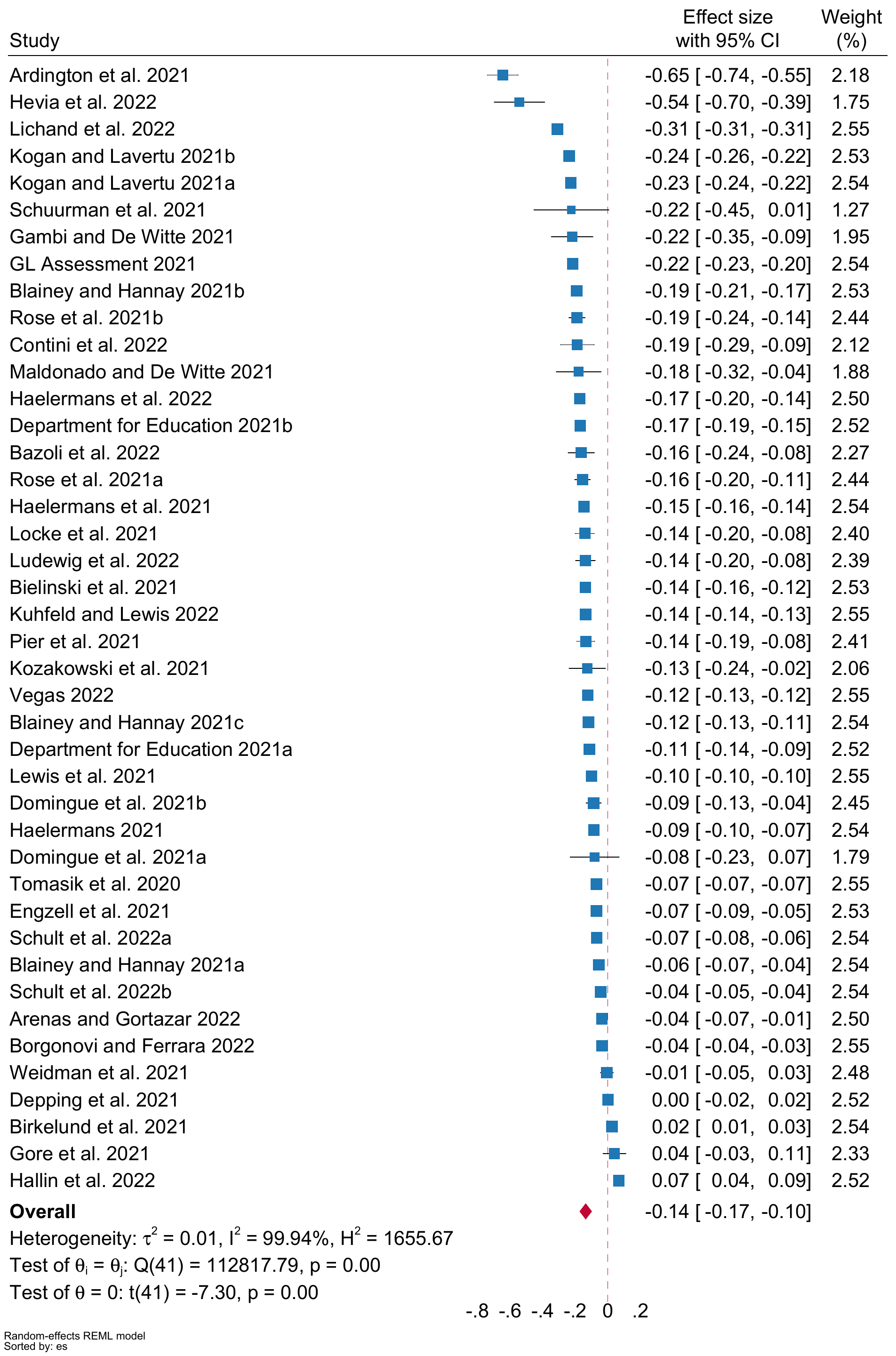}}
    \hfill
    \bigskip
    \begin{footnotesize}
    \newline
        \emph{Notes:} The figure displays a forest plot of 42 included studies. The effects are expressed as Cohen's \textit{d} weighted by the inverse of variance using the random effects model. (a) shows the original, and (b) is the reproduction. We reproduced the Blainey and Hannay 2021a effect size as $-0.06$, while the original article reports $-0.05$. The confidence intervals are the same, possibly due to rounding.   
    \end{footnotesize}
\end{figure}

\restoregeometry

\begin{figure}[htbp]
    \caption{Estimates of COVID-19 learning deficits in time}
    \begin{center}
      \subfigure[Original Figure 4]{\includegraphics[width=0.85\textwidth]{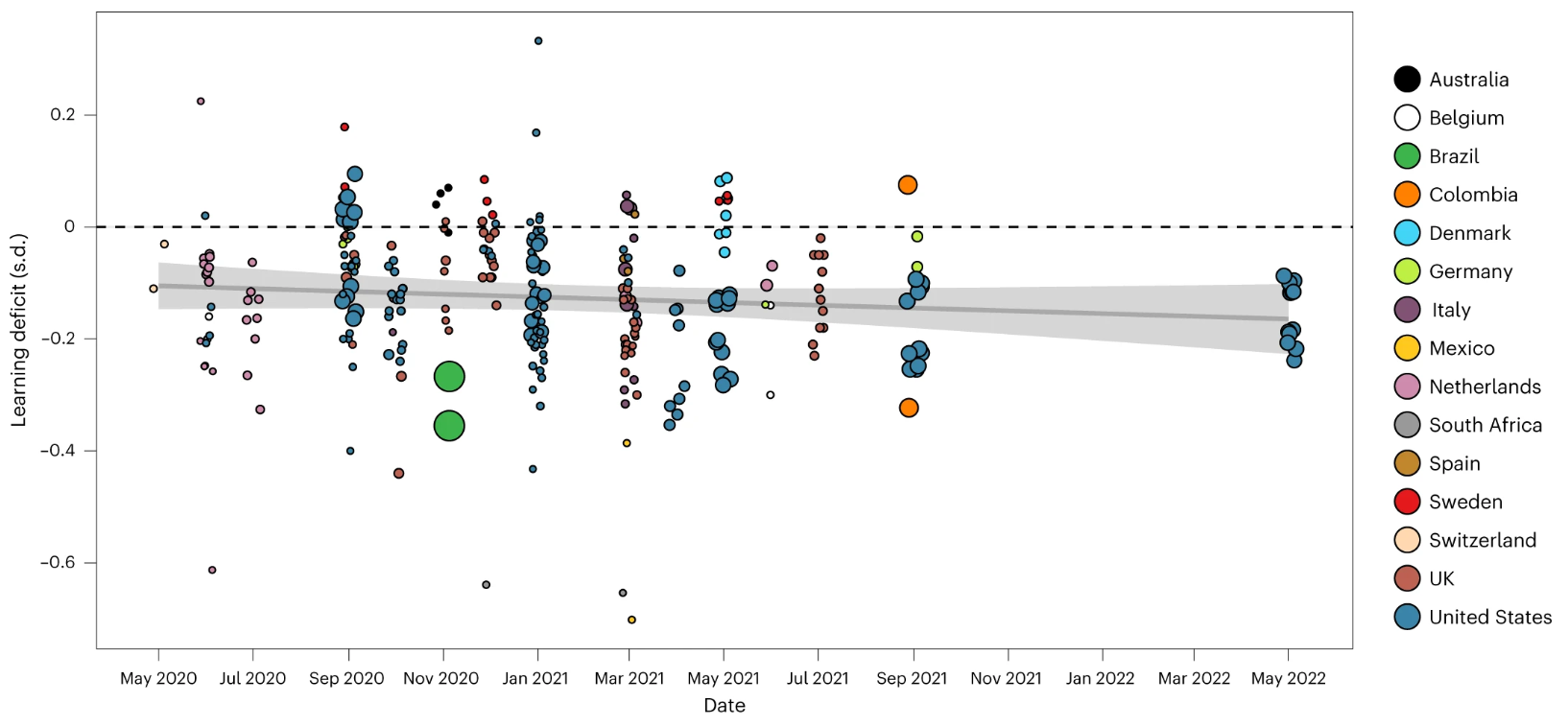}} 
    \centering
    \newline
    \subfigure[Reproduction of Figure 4]{\includegraphics[width=0.91\textwidth]{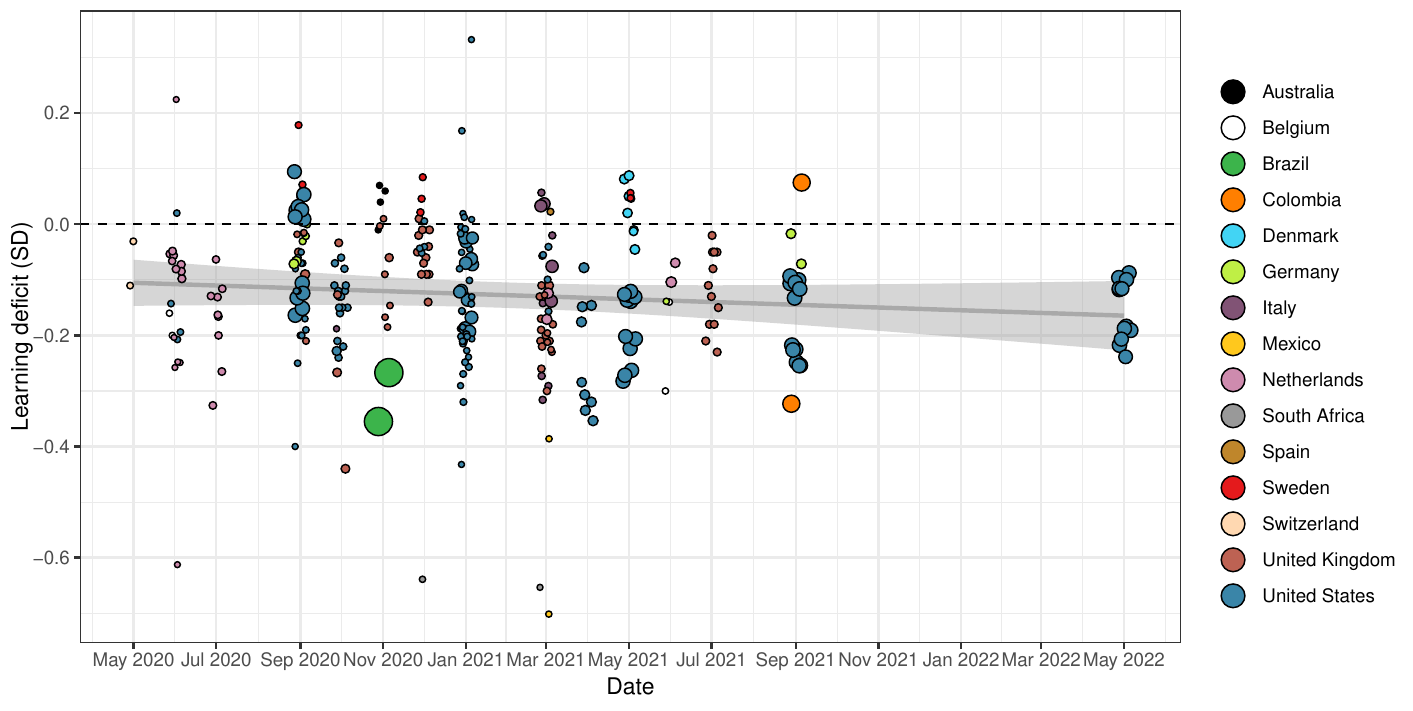}}
    \hfill  
    \end{center}    
    \bigskip
    \begin{footnotesize}
        \emph{Notes:} The figure displays estimates of COVID-19 learning deficit. The horizontal axis shows the time of the estimate, and the vertical axis presents the estimates expressed as Cohen's \textit{d}. Countries are in color scale. The slope coefficient of a trend line estimated using OLS with standard errors clustered at the study level is not statistically different from 0. (a) shows the original, and (b) is the reproduction. See \autoref{tab:timeslope} for details. 
    \end{footnotesize}
\end{figure}

\begin{figure}[htbp]
    \caption{Variation in estimates of COVID-19 learning deficits}
    \begin{center}
      \subfigure[Original of Figure 6]{\includegraphics[width=1\textwidth]{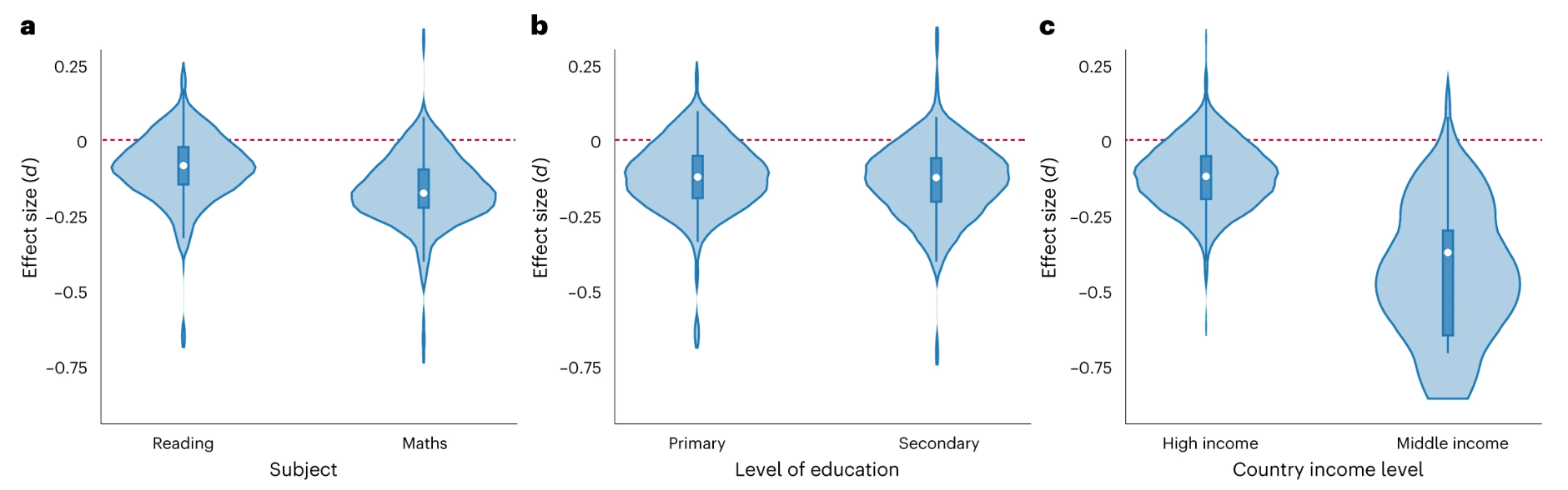}} \label{fig: variations}
    \centering
    \newline
    \subfigure[Reproduction Figure 6]{\includegraphics[width=1\textwidth]{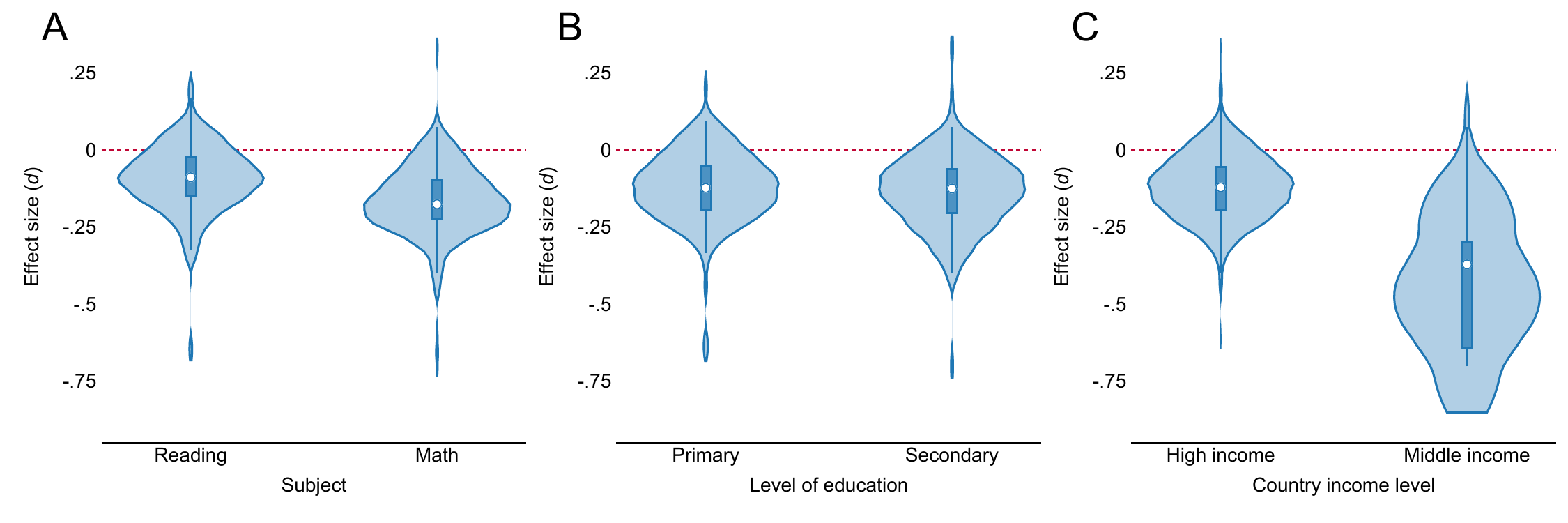}}
    \hfill  
    \end{center}    
    \bigskip
    \begin{footnotesize}
        \emph{Notes:} The figure displays variation in estimates of COVID-19 learning deficit for school subjects (mathematics and reading), level of education, and socio-economic inequality. (a) shows the original, and (b) is the reproduction. No differences between the two. See \autoref{tab:moderators} for details.
    \end{footnotesize}
\end{figure}
\clearpage
\subsection{Tables}\label{sec: app_tables}
\renewcommand{\thetable}{A\arabic{table}}
\setcounter{table}{0}
\begin{footnotesize}
    \begin{table}[htbp]
\centering
\caption{Replication Package Contents and Reproducibility}
\label{tab:reppkg_table}
\begin{threeparttable}
    \footnotesize 
	\begin{tabular}{lccc}
	\toprule 
	\textbf{Replication Package Item} 	& \textbf{Fully} & \textbf{Partial} &           \textbf{No} \\
	\midrule
	Raw data provided                 	& \checkmark &  &  \\ 
	Analysis data provided				& \textcolor{black}{\checkmark} &  & \\ [0.3cm]
	Cleaning code provided				& \checkmark &  &  \\ 
	Analysis code provided				& \checkmark &  & \\ [0.3cm]
	Reproducible from raw data			&  & \checkmark &  \\ 
	Reproducible from analysis data		&  & \checkmark &  \\ [0.3cm]
	\bottomrule
	\end{tabular}
\begin{tablenotes}[flushleft]
      \footnotesize
      \item[] \emph{Notes}: This table summarises the replication package contents contained in \cite{betthauser2023systematic}. 
\end{tablenotes}
\end{threeparttable}
\end{table}
 
    \begin{table}[htbp]
\begin{center}
    \caption[Estimates of learning deficits in time]{Estimates of learning deficits in time}\label{tab:timeslope}
\footnotesize
\begin{threeparttable}
\begin{tabular}{@{}lcc@{}}
\toprule
 &  \textbf{Original Study} & \textbf{Reproduction}\\
\midrule
Slope coefficient: $\beta_{months}$ & $-0.00$ & $-0.00$\\
\hspace{3mm} $p$-value& $0.097$& $0.097$\\ 
\hspace{3mm} 95\% CI & [$-0.01$, $0.00$]& [$-0.01$, $0.00$]\\ [0.3cm]   
\midrule
Observations & $291$& $291$\\ 
Clusters & $42$& $42$\\
\bottomrule 
\end{tabular} 
\begin{tablenotes}[flushleft]
      \footnotesize
      \item[] \emph{Notes: } The table shows the comparison of the original and reproduced estimate of the slope coefficient obtained by regressing the estimates on months in which learning was measured. Standard errors are clustered at the study level. We report $p$-values and 95\% confidence intervals (CI). 
    \end{tablenotes} 
\end{threeparttable} 
\end{center}
\end{table}

    \begin{table}[htbp]
\begin{center}
    \caption[Variation in estimates of learning deficits]{Variation in estimates of learning deficits}\label{tab:moderators}
\footnotesize
\begin{threeparttable}
\begin{tabular}{@{}lcc@{}}
\toprule
 &  \textbf{Original Study} & \textbf{Reproduction}\\
\midrule
\textit{School subject} \\
\hspace{3mm} Reading & $-0.09$ & $-0.09$ \\ 
\hspace{3mm} IQR & [$-0.15$, $-0.02$] &  [$-0.15$, $-0.02$] \\ [0.3cm]
\hspace{3mm} Mathematics & $-0.18$ & $-0.18$\\    
\hspace{3mm} IQR & [$-0.23$, $-0.09$] & [$-0.23$, $-0.09$] \\ [0.3cm]
\hspace{3mm} Mean difference & $-0.07$*** & $-0.07$*** \\ 
\hspace{3mm} $p$-value  &  $   0.000$&  $  0.000$ \\
&[-0.11, -0.04]& [$-0.11$, $-0.04$]\\ [0.3cm]
\midrule
\textit{Level of education} \\
\hspace{3mm} Primary & $-0.12$ & $-0.12$ \\ 
\hspace{3mm} IQR &[$-0.19$, $-0.05$]& [$-0.19$, $-0.05$]\\ [0.3cm]
\hspace{3mm} Secondary & $-0.12$ & $-0.12$\\    
\hspace{3mm} IQR & [$-0.21$, $-0.06$] & [$-0.21$, $-0.06$] \\ [0.3cm]
\hspace{3mm} Mean difference & $-0.01$ & $-0.01$ \\ 
\hspace{3mm} $p$-value &$  0.556$& $  0.556$ \\
&[$-0.06$, $0.03$]& [$-0.06$, $0.03$]\\ [0.3cm]
\midrule
\textit{Country income level} \\
\hspace{3mm} High & $-0.12$ & $-0.12$ \\ 
\hspace{3mm} IQR &[$-0.20$, $-0.05$]& [$-0.20$, $-0.05$]\\ [0.3cm]
\hspace{3mm} Middle & $-0.37$ & $-0.37$ \\    
\hspace{3mm} IQR & [$-0.65$, $-0.30$] &  [$-0.65$, $-0.30$]\\ [0.3cm]
\hspace{3mm} Mean difference & $-0.29$*** & $-0.29$*** \\ 
\hspace{3mm} $p$-value & $  0.008$&$ 0.008$ \\
&[$-0.50$, $-0.08$]& [$-0.50$, $-0.08$]\\ [0.3cm]
\bottomrule 
\end{tabular} 
\begin{tablenotes}[flushleft]
      \footnotesize
      \item[] \emph{Notes: } The table shows the comparison of the original and reproduced median learning deficit for school subjects, level of education, and country income level. IQR = Interquartile range as in the original paper. Significant at $^{***}[1\%]$, $^{**}[5\%]$, $^{*}[10\%]$ level.
    \end{tablenotes} 
\end{threeparttable} 
\end{center}
\end{table}

\end{footnotesize}

\clearpage
\section{RTMA Implementation with Negative Affirmative  Results}
\label{app:rtma}

\doublespacing
RTMA, as implemented in \texttt{phacking\_meta} , assumes that 
$p$-hacking favors positive and significant results 
(\texttt{favor\_positive = TRUE}). In our setting, affirmative 
results are negative and statistically significant since learning 
deficiency is measured as a negative value. We therefore apply RTMA 
to sign-reversed data ($y_i^f = -y_i$) with \texttt{favor\_positive 
= TRUE} and reverse the sign of the resulting mean estimate.

We confirmed that this approach is equivalent to setting 
\texttt{favor\_positive = FALSE} with the original data, as both 
specifications produce identical results. However, upon thorough 
examination of \texttt{phacking\_meta}, \texttt{multibias\_meta}, 
and \texttt{pubbias\_meta}, we found that \texttt{favor\_positive = 
FALSE} in \texttt{phacking\_meta} effectively reverses the sign of 
the dataset internally before proceeding, producing estimates with 
inverted signs --- which is identical to our manual sign-reversal 
approach. This behavior also affects \texttt{multibias\_meta}. The 
function \texttt{pubbias\_meta} does not suffer from this issue. We 
were able to correct this in \texttt{multibias\_meta} but not in 
\texttt{phacking\_meta}. We have reported this problem at 
\url{https://github.com/mathurlabstanford/metabias-apps/issues/1}, 
where a detailed description is available.

\clearpage
\section{Caliper Test and P-Curve Analysis}
\label{app:pcaliper}

To formally test for $p$-hacking, we apply a caliper test following \citet{gerber2008statistical} and \citet{brodeur2016star}, which examines whether there is excess mass just inside the significance threshold relative to just outside it. Under $p$-hacking, researchers manipulate borderline non-significant estimates to cross the threshold, producing a spike just inside $z = -1.96$ and a corresponding gap just outside. 

Across caliper widths from $0.05$ to $0.50$ and across the thresholds $t=0$, $t=-1.96$, and $t=-2.58$, the caliper tests show little consistent evidence of bunching around reporting thresholds (Table~\ref{tab:caliper}, Figure~\ref{fig:caliper}). Because the relevant significance thresholds are negative, estimates below $t=-1.96$ are more negative and therefore lie on the statistically significant side of the threshold. At the standard caliper width of $0.10$, at the $t=-1.96$ threshold is negative ($-0.167$, SE $=0.333$, $N=3$), indicating slightly more mass on the less negative, non-significant side of the threshold, opposite to what selective reporting of negative learning-loss estimates would predict. Only at wider calipers of $0.40$ and $0.45$ does the statistic become weakly significant at the 10\% level. Overall, the caliper tests provide at most weak and non-robust evidence of bunching near the conventional significance cutoff.



We also attempted a $p$-curve analysis following \citet{simonsohn2014p}, which tests whether the distribution of significant $p$-values is right-skewed as expected under genuine effects. However, the $p$-curve is structurally uninformative in this dataset. Sixty-one $p$-values underflow numerically to zero, and a further $\sim$175 produce $p < .001$, as a direct consequence of the extremely large sample sizes in the literature (median $n = 50{,}000$; max $n = 10{,}884{,}922$). Even modest effect sizes --- such as $d = -0.27$ --- generate $z$-scores as large as $|z| = 1{,}775$ when standard errors are of order $0.0002$, collapsing virtually the entire $p$-value distribution into a narrow band indistinguishable from zero. The $p$-value distribution therefore carries no information about selective reporting, and we do not report formal $p$-curve test statistics (Figure~\ref{fig:pcurve}).\footnote{This is a structural feature of the literature rather than a flaw in the analysis. Commercial platform studies with millions of observations have effectively infinite statistical power, making significance-threshold manipulation both unnecessary and undetectable through $p$-value-based diagnostics.}
\clearpage
\subsection{Figures}
\setcounter{figure}{0}
\renewcommand{\thefigure}{C\arabic{figure}}

\begin{figure}[ht]
\caption{Caliper test for excess bunching around the significance threshold $z = -1.96$.} \label{fig:caliper}
    \begin{center}
        \includegraphics[width=0.75\textwidth]{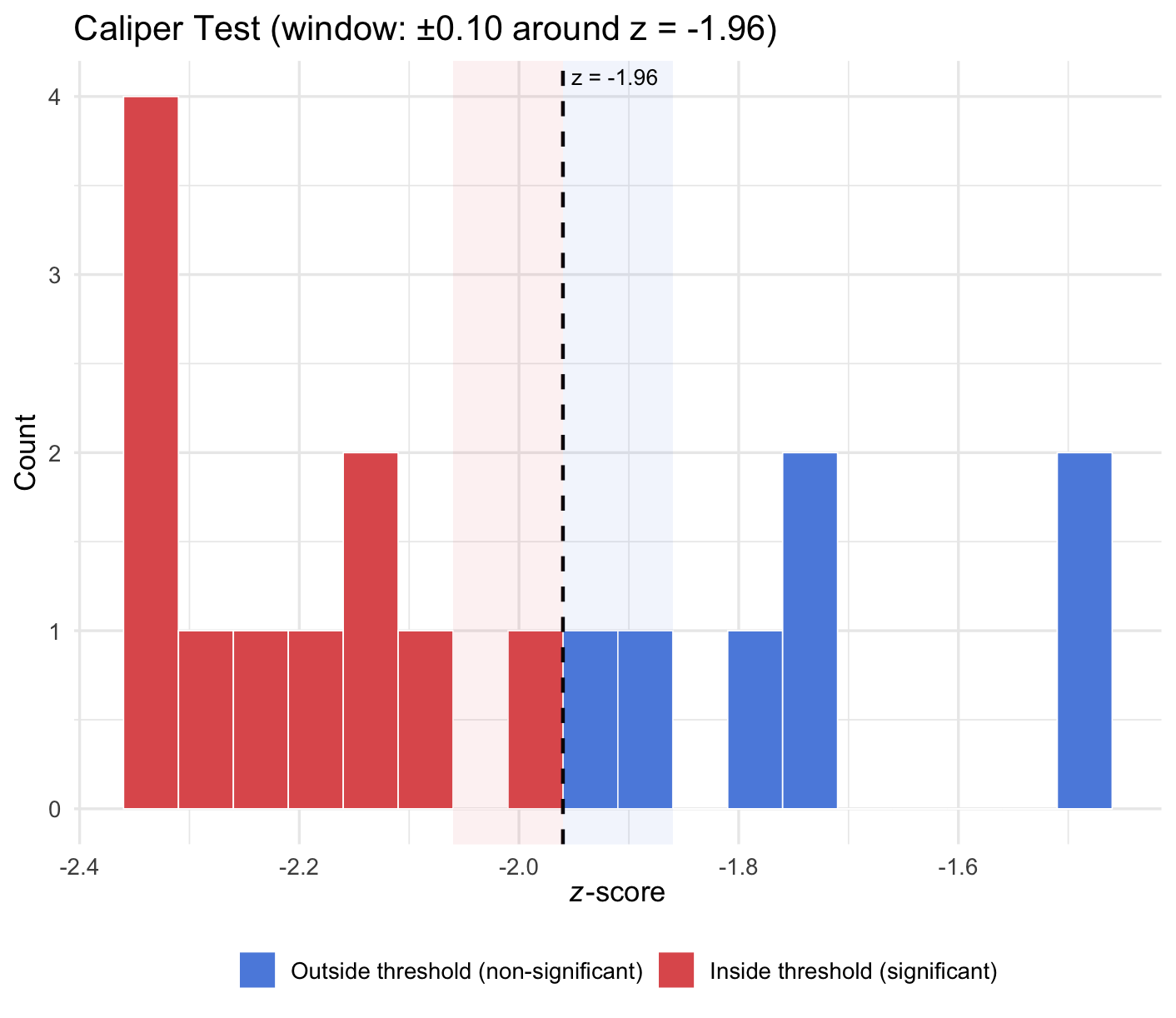}
    \end{center}
\begin{footnotesize}
        \emph{Notes:} Shaded regions indicate the $\pm 0.10$ caliper window. Red bars denote estimates inside the threshold (significant); blue bars denote estimates outside (non-significant).
\end{footnotesize}
\end{figure}

\begin{figure}[ht]
     \caption{$P$-curve distribution.}\label{fig:pcurve}
\begin{center}
     \includegraphics[width=0.75\textwidth]{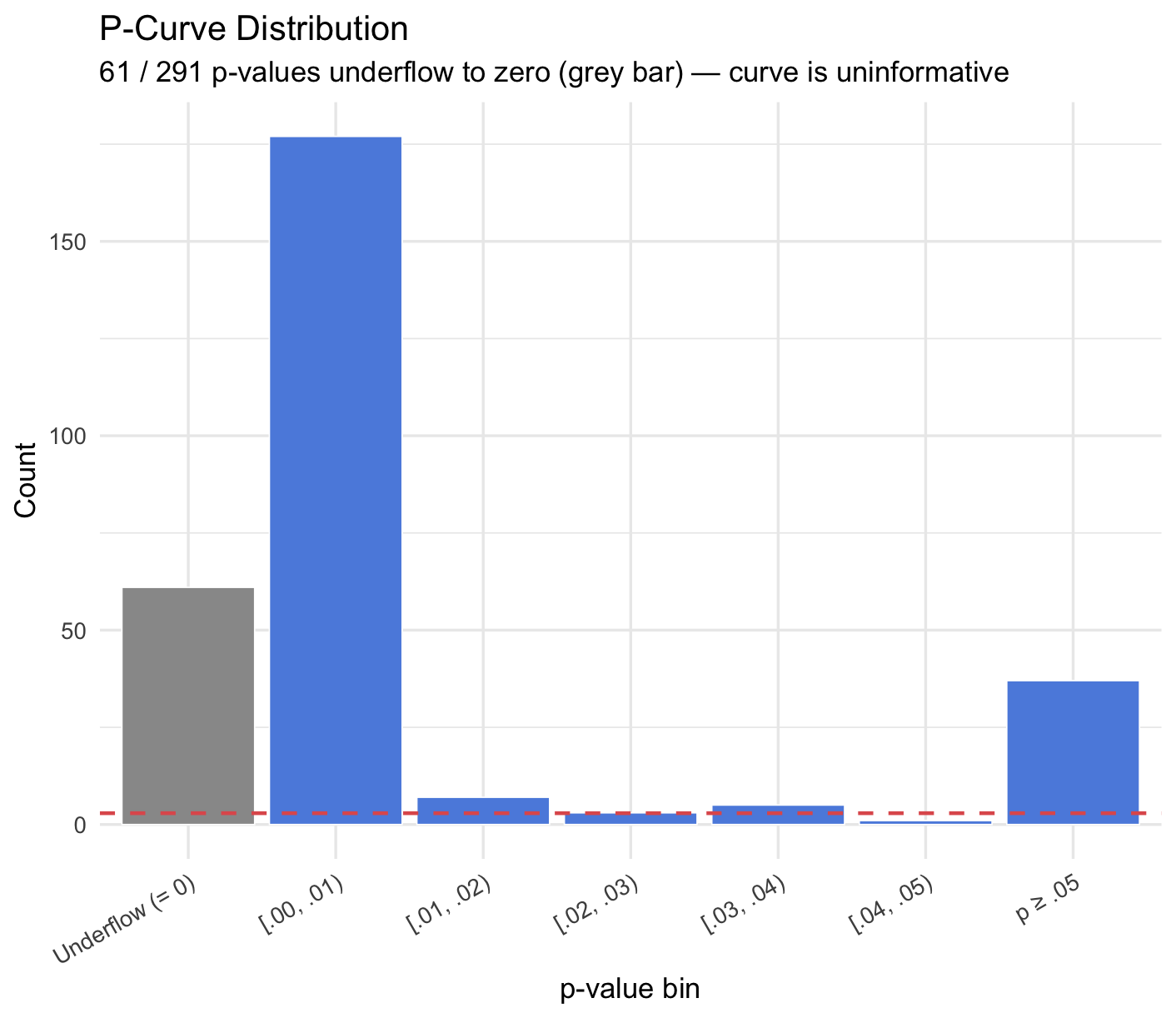}
\end{center}
\begin{footnotesize}
        \emph{Notes:} The grey bar denotes $p$-values that underflow numerically to zero ($n = 61$). The dominance of the $[.00, .01)$ bin reflects the extremely large sample sizes in the literature rather than the evidential strength of individual studies.
\end{footnotesize}    
\end{figure}

\clearpage
\subsection{Tables}

\setcounter{table}{0} \renewcommand{\thetable}{C\arabic{table}}

\begin{table}[h!]
\centering
\small
\caption{Caliper tests for selection around significance thresholds}
\label{tab:caliper}
\begin{threeparttable}
\begin{tabular}{lccc} 
\toprule
 & t-statistic $= 0$ & t-statistic $= -1.96$ & t-statistic $= -2.58$ \\
\hline
Caliper .05 & . & 0.000 & . \\
   & & (0.500) & \\
   & N $= $ 1 & N $=$ 2 & N $= $ 1 \\
[6pt]
Caliper .1 & 0.000 & -0.167 & -0.167 \\
   & (0.500) & (0.333) & (0.333) \\
   & N $=$ 2 & N $=$ 3 & N $=$ 3 \\
[6pt]
Caliper .15 & 0.000 & 0.000 & -0.167 \\
   & (0.289) & (0.289) & (0.333) \\
   & N $=$ 4 & N $=$ 4 & N $=$ 3 \\
[6pt]
Caliper .2 & -0.167 & 0.071 & 0.000 \\
   & (0.203) & (0.202) & (0.289) \\
   & N $=$ 6 & N $=$ 7 & N $=$ 4 \\
[6pt]
Caliper .25 & -0.167 & 0.000 & -0.250 \\
   & (0.203) & (0.155) & (0.143) \\
   & N $=$ 6 & N $=$ 10 & N $=$ 8 \\
[6pt]
Caliper .3 & -0.071 & 0.045 & -0.250 \\
   & (0.202) & (0.138) & (0.143) \\
   & N $=$ 7 & N $=$ 11 & N $=$ 8 \\
[6pt]
Caliper .35 & -0.045 & 0.083 & -0.167 \\
   & (0.141) & (0.138) & (0.110) \\
   & N $=$ 11 & N $=$ 12 & N $=$ 12 \\
[6pt]
Caliper .4 & -0.083 & 0.188$^{*}$ & -0.062 \\
   & (0.116) & (0.085) & (0.096) \\
   & N $=$ 12 & N $=$ 16 & N $=$ 16 \\
[6pt]
Caliper .45 & -0.083 & 0.188$^{*}$ & -0.062 \\
   & (0.116) & (0.085) & (0.096) \\
   & N $=$ 12 & N $=$ 16 & N $=$ 16 \\
[6pt]
Caliper .5 & -0.083 & 0.111 & -0.100 \\
   & (0.116) & (0.108) & (0.110) \\
   & N $=$ 12 & N $=$ 18 & N $=$ 20 \\
\bottomrule
\end{tabular}
\begin{tablenotes}[flushleft]
\footnotesize
      \item[] \emph{Notes: } 
The table reports results for caliper tests \citep{gerber2008statistical}. 
The tests compare the relative frequency of estimates above and below 
important thresholds for the t-statistic; the rows show results for
different caliper widths. A test statistic of $-0.167$, with $N=3$ at the $-1.96$ threshold, means that $66.7\%$ or 2 estimates are above the threshold and 1 estimate ($33.3\%$) is below. \autoref{fig:caliper} shows this graphically. N $=$ number of estimates. 
Standard errors are reported in parentheses and clustered at the study level. 
Significant at $^{***}[1\%]$, $^{**}[5\%]$, $^{*}[10\%]$ level.
\end{tablenotes}

\end{threeparttable}
\end{table}




\section{Compliance with Meta-Analysis Guidelines}
\label{app:guidelines}
\setcounter{table}{0}
\renewcommand{\thetable}{D\arabic{table}}


This appendix documents our adherence to the 18-item checklist for modern meta-analysis in economics proposed by \citet{irsova2024meta}. Because this paper reanalyzes an existing, publicly available dataset rather than conducting an independent literature search, several items in the checklist are not applicable by design (items~3--5, 9). Items~15--18, which concern multiple meta-regression and conditional implied estimates, are unfulfilled by design: the paper's contribution is confined to bias correction and reproducibility verification.

\begingroup
\setlength{\tabcolsep}{4pt}
\begin{ThreePartTable}
 \small  
\begin{TableNotes}[flushleft]
\footnotesize
\item \textit{Notes:} $\checkmark$ = satisfied; 
$\checkmark^{*}$ = substantially satisfied with acknowledged limitations; $\times$ = not satisfied; N/A = not applicable 
given the paper's design.
\end{TableNotes}


\begin{longtable}{>{\centering}p{0.02\linewidth}
                  p{0.3\linewidth}
                  >{\centering}p{0.08\linewidth}
                  p{0.52\linewidth}}

\caption{Compliance with the \citet{irsova2024meta} 
 meta-analysis checklist}
\label{tab:checklist}\\
\toprule
\textbf{$N$ } & \textbf{Requirement} & \textbf{Status} 
& \textbf{Notes} \\
\midrule
\endfirsthead
\multicolumn{4}{l}{\footnotesize\textit{Table~\ref{tab:checklist} 
continued}}\\
\toprule
\textbf{$N$} & \textbf{Requirement} & \textbf{Status} 
& \textbf{Notes} \\
\midrule
\endhead
\midrule
\multicolumn{4}{r}{\footnotesize\textit{Continued on next page}}\\
\endfoot
\bottomrule
\insertTableNotes
\endlastfoot

1 & Topic known from own primary research & $\checkmark$ & Co-authors have published primary and meta-analytic research on publication bias, $p$-hacking, and the economics of education. \\[4pt]

2 & New meta-analysis justified by stronger methods 
 & $\checkmark$ 
 & \citet{betthauser2023systematic} do not apply formal bias-correction methods. Our \textit{raison d'\^{e}tre} is the application of PET-PEESE, 3PSM, RoBMA, MAIVE, RTMA, and multi-bias sensitivity analysis to their dataset. \\[4pt]

3 & Google Scholar search, first 500 hits inspected 
 & N/A 
 & We build on the dataset of \citet{betthauser2023systematic} rather than conducting an independent literature search. Readers are referred to their Figure~1 and PROSPERO registration for the original PRISMA diagram (Appendix~\ref{app:reproduction}).\\[4pt]

4 & Snowballing: 30 most-cited studies inspected 
 & N/A 
 & See item~3. \\[4pt]

5 & No study excluded \textit{a priori} on quality grounds 
 & N/A 
 & See item~3. Study inclusion follows \citet{betthauser2023systematic}. \\[4pt]

6 & All estimates and standard errors collected 
 & $\checkmark$ 
 & All 291 effect-size estimates from 42 studies are included, as reported in \citet{betthauser2023systematic}. \\[4pt]

7 & Data collected independently by two co-authors 
 & $\checkmark^{*}$ 
 & The dataset is inherited from \citet{betthauser2023systematic} and publicly available; no independent re-coding of the primary effect-size estimates was performed. \\[4pt]

8 & Original effect-size measures used when comparable 
 & $\checkmark$ 
 & All estimates are standardized mean differences (Cohen's~$d$), the measure used in the original literature. \\[4pt]

9 & Partial correlations used only as last resort 
 & N/A 
 & Cohen's~$d$ is used throughout; partial correlations are not employed. \\[4pt]

10 & Outliers and influence points inspected 
 & $\checkmark$ 
 & DFBETAS diagnostics are reported in Figure \ref{fig:dfbetas}. Eleven influential observations are identified; excluding all eleven moves the pooled estimate from $-0.124$ to $-0.116$. \\[4pt]

11 & At least 10 heterogeneity variables coded 
 & $\times$ 
 & The paper's scope is confined to bias correction. Table~\ref{tab:subgroups} reports subgroup means by subject, school level, grade group, sample size, country, and source tier as descriptive context. Systematic heterogeneity analysis across moderator variables is available in \citet{betthauser2023systematic}, who examine variation by subject, school level, socio-economic background, and country income level. A full multiple meta-regression is beyond the scope of the present paper. \\[4pt]

12 & Simple summary statistic uses UWLS rather than FE/RE 
 & $\checkmark^{*}$ 
 & The UWLS estimate \citep{stanley2023unrestricted, stanley2017neither} yields the same inverse-variance weighted point estimate as the fixed-effect mean shown by the black diamond in the significance funnel plot ($\hat{\mu}=-0.245$). However, inference is based on a considerably more conservative cluster-robust standard error clustered at the study level (SE $=0.051$), rather than the conventional fixed-effect standard error, which is implausibly small in this setting given the extreme heterogeneity ($I^2=99.97\%$). Panel~A also reports unweighted and RVE means as uncorrected benchmarks, consistent with the paper's focus on bias correction rather than summary estimation.  See footnote~\ref{fn:uwls} in Section~\ref{sec:pubbias} for further discussion. \\[4pt]

13 & Publication bias corrected using methods from both model families 
 & $\checkmark$ 
 &Selection models: 3PSM and RoBMA, which averages across a wider class of selection models weighted by data fit and parsimony. Funnel-based models: PET-PEESE and MAIVE, the latter addressing the specific endogeneity problem arising from Cohen's~$d$ standardization and potential $p$-hacking. Additional sensitivity methods: MAN, selection-ratio adjustments, $s$-value, RTMA, and the multi-bias framework \citep{mathur2024assessing, mathur2024p, mathur2024internal}. 
     \\[4pt]

14 & Standard errors clustered at study level; wild bootstrap if $<40$ studies 
 & $\checkmark$ 
 & Standard errors are clustered at the study level throughout (42 clusters). Wild bootstrap is recommended for fewer than 40 clusters and is not required here. \\[4pt]

15 & Study-level dummies used in meta-regressions 
 & $\times$ 
 & Study-level fixed effects are not included in the main specifications. The principal estimators (PET-PEESE, MAIVE, 3PSM, RoBMA) do not incorporate study dummies; doing so would eliminate between-study variation that is the primary object of interest in a bias-correction exercise. \\[4pt]

16 & Multiple MRA estimated by BMA with dilution prior 
 & $\times$ 
 & A full multiple meta-regression with BMA is not conducted. The paper's contribution is confined to bias correction and sensitivity analysis.\\ 

17 & Frequentist model averaging or general-to-specific as robustness check 
 & $\times$ 
 & See item~16. \\[4pt]

18 & Conditional means provided for different scenarios 
 & $\times$ 
 & Conditional implied estimates are not provided, consistent with the paper's focus on the mean bias-corrected effect rather than heterogeneity decomposition. Subgroup means by key moderators are reported in Table \ref{tab:subgroups}. \\
\end{longtable}
\end{ThreePartTable}
\endgroup

\clearpage

\subsection*{Reporting Standards \citep{havranek2020reporting, 
cook2026reporting, cook2026guidance}} 

The reporting standards of \citet{havranek2020reporting}, 
updated for AI use by \citet{cook2026reporting} following 
the guiding principles of \citet{cook2026guidance}, are 
addressed as follows. The research question and effect size measure (Cohen's~$d$) are defined in Section~\ref{sec:data}. Study inclusion criteria and the PRISMA flow diagram follow \citet{betthauser2023systematic}, to whose replication package we refer readers (Appendix~\ref{app:reproduction}). Descriptive statistics are reported in Tables~\ref{tab:dscrstat} and~\ref{tab:subgroups}. Publication, selection, and misspecification biases are the central subject of Sections~\ref{sec:pubbias} and~\ref{sec:phacking}; bias-corrected estimates are reported in Tables~\ref{tab:pubbias} and~\ref{tab:phack}. Within-study dependence is addressed via study-level clustering throughout. Effect sizes are displayed in Figures~\ref{fig:country}--\ref{fig:sig_funnel}. Influence and leverage are assessed via DFBETAS (Figure~\ref{fig:dfbetas}); the sensitivity of the pooled estimate to excluding influential observations, to alternative correction methods, and to varying degrees of selective publication is documented across Sections~\ref{sec:pubbias} and~\ref{sec:phacking}. Economic significance is discussed using the benchmark of $0.40$~SD per school year \citep{bloom2008performance}. The underlying dataset is available at \url{https://doi.org/10.17605/osf.io/u8gaz}; replication code for all analyses is available at \url{https://meta-analysis.cz}. AI use is disclosed in the Use of Generative AI statement following the Data Availability Statement.

\end{document}